\documentclass[nofootinbib,
tightenlines,
superscriptaddress,11pt]{revtex4}

\usepackage{graphicx}
\usepackage{amsmath,amssymb,amsfonts,amsthm,stmaryrd,mathtools,bm,physics}
\usepackage{color}
\usepackage{tikz}
\usepackage[normalem]{ulem}
\usepackage{braket}
\allowdisplaybreaks[1]

\def\be#1\ee{\begin{align}#1\end{align}}

\def\ba{\begin{eqnarray}}
\def\ea{\end{eqnarray}}
\def\nn{\nonumber}
\def\q{\quad}

\usepackage[bookmarks,linktocpage, colorlinks=true, plainpages = false, citecolor = treegreen,  linkcolor=darkblue, urlcolor = darkblue, filecolor = blue]{hyperref} 

\definecolor{darkblue}{rgb}{0., 0.4, 0.8}
\definecolor{treegreen}{rgb}{0., 0.7, 0.3}

\begin{document}

\title{Spikes and spines in 3D Lorentzian simplicial quantum gravity}

\author{Johanna Borissova}
\email{jborissova@perimeterinstitute.ca} 
\affiliation{Perimeter Institute, 31 Caroline Street North, Waterloo, ON, N2L 2Y5, Canada}
\affiliation{Department of Physics and  Astronomy, University of Waterloo, 200 University Avenue West, Waterloo, ON, N2L 3G1, Canada}
\author{Bianca Dittrich}
\email{bdittrich@perimeterinstitute.ca} 
\affiliation{Perimeter Institute, 31 Caroline Street North, Waterloo, ON, N2L 2Y5, Canada}
\affiliation{Theoretical Sciences Visiting Program, Okinawa Institute of Science and Technology Graduate University, Onna, 904-0495, Japan}
\author{Dongxue Qu}
\email{dqu@perimeterinstitute.ca} 
\affiliation{Perimeter Institute, 31 Caroline Street North, Waterloo, ON, N2L 2Y5, Canada}
\author{Marc Schiffer \footnote{corresponding author}}
\email{mschiffer@perimeterinstitute.ca} 
\affiliation{Perimeter Institute, 31 Caroline Street North, Waterloo, ON, N2L 2Y5, Canada}

\begin{abstract}

Simplicial approaches to quantum gravity such as Quantum Regge Calculus and Spin Foams include configurations where bulk edges can become arbitrarily large while keeping the lengths of the boundary edges small. Such configurations pose significant challenges in Euclidean Quantum Regge Calculus, as they lead to infinities for the partition function and length expectation values. 
Here we investigate such configurations in three-dimensional Lorentzian Quantum Regge Calculus, and find that the partition function and length expectation values remain finite. This shows that the Lorentzian approach can avoid a key issue of the Euclidean approach. We also find that the space of configurations, for which bulk edges can become very large, is much richer than in the Euclidean case. In particular, it includes configurations with irregular light-cone structures, which lead to imaginary terms in the Regge action and branch cuts along the Lorentzian path integral contour. Hence, to meaningfully define the Lorentzian Regge path integral, one needs to clarify how such configurations should be handled.
\end{abstract}

\maketitle
\tableofcontents

\section{Introduction}\label{Sec:Introduction}

The path integral for quantum gravity requires many choices to be made~\cite{deBoer:2022zka}, such as specifying the space of geometries to be summed over, or the regularization of the path integral. One such regularization is provided by Quantum Regge calculus~\cite{Regge:1961px,Regge:2000wu,Hamber:2009zz}, where the space of geometries is given by piecewise flat~\footnote{One can also choose piecewise homogeneous geometries~\cite{Bahr:2009qd}.} geometries, constructed via triangulations. The geometry of these triangulations is uniquely specified by assigning  lengths to all edges of the triangulation. The Regge action~\cite{Regge:1961px} is a discretization of the Einstein-Hilbert action based on piecewise linear and flat geometries. This regularizes the infinite-dimensional path integral and reduces it to an integral over finitely many edge lengths.

\begin{figure}[h]
	\centering
	\includegraphics[width=.7\textwidth]{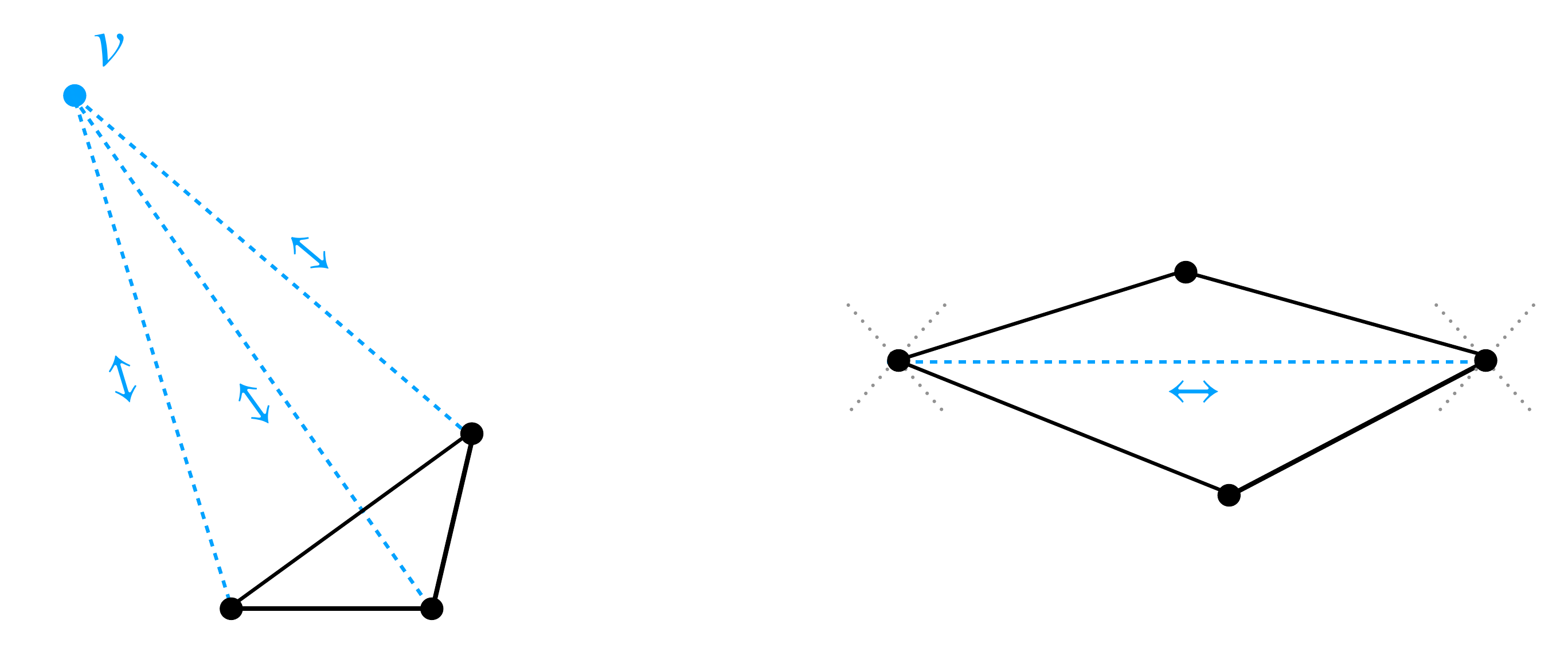}
	\hfill
	\caption{\label{Fig:SpikeSpine} 
		Left: Illustration of a spike configuration in $d=2$ Lorentzian spacetime. The boundary complex is given by the three black edges whose length is kept fixed. The blue vertex $v$ and the dashed blue edges are in the bulk. The bulk edges can become arbitrarily large. Right: Illustration of a spine configuration in $d=2$ Lorentzian spacetime. The boundary complex is given by the four black edges whose lengths are kept fixed. The dashed blue line represents a bulk edge which can become arbitrarily large. This is possible in Lorentzian signature, as one can tilt the boundary edges arbitrarily close towards the light rays represented by the dotted gray lines.}
\end{figure}

This does not  guarantee the finiteness of the path integral, however: There are configurations where edge lengths can~\footnote{The choice of edge lengths is restricted by the (Euclidean or Lorentzian) generalized triangle inequalities. These triangle inequalities guarantee that any top-dimensional simplex of the triangulation can be embedded into (Euclidean or Lorentzian) flat space.} become arbitrarily large. A class of such configurations are called spikes, which are defined as follows, see also the left panel of Fig.~\ref{Fig:SpikeSpine}: consider a bulk vertex $v$ and the set of all simplices containing this vertex, i.e. the star of $v$. We fix the length of the edges in the boundary of this set to some finite values, thus also fixing a finite value for the volume of this boundary. Spikes are configurations where the length of the edges sharing the vertex $v$ can become arbitrarily large. The left panel of Fig.~\ref{Fig:SpikeSpine} shows an illustration of a spike configuration in two spacetime dimensions.

Since those configurations are a-priori included in the Regge path integral, its finiteness is not guaranteed.
In fact, in $d>2$ dimensions~\footnote{In $d=2$ dimensions the gravitational action (with vanishing cosmological constant) is a topological invariant. Nevertheless, spike configurations turn out to be highly problematic~\cite{Ambjorn:1997ub}.}, such spike configurations also include the conformal mode~\cite{Dittrich:2011vz}, 
which contributes with the ``wrong" sign to the gravitational action~\cite{Gibbons:1978ac}, 
thereby rendering it unbounded from below. 
The path integral for Euclidean quantum gravity, with amplitudes $\exp(-S_E)$, 
evaluated on such configurations therefore leads to infinities~\footnote{Adding a positive cosmological constant does not cure this issue in all cases. \cite{Dittrich:2021gww} constructs simple examples in 4D where the Euclidean action goes to $-\infty$ in the limit of infinitely large edge lengths, as long as the boundary volume is below a critical value which scales as $\Lambda^{-3/2}$.}~\cite{Dittrich:2011vz}. 

One main aim of this work is to study the convergence properties of spike configurations (as well as spine configurations, defined below) in the Lorentzian Regge path integral.  In this work we will focus on the three-dimensional case, the four-dimensional case will be discussed in~\cite{Toappear}. Previous work focused on the two-dimensional case~\cite{Tate:2011ct,Jia:2021deb,Ito:2022ycc} or resorted to path integral measures which (exponentially) suppress large edge lengths~\cite{Mikovic:2023tgg}. We will however see that a suppressing path integral measure is  not necessary in order to obtain finite expectation values in Lorentzian spacetimes.  We even find that for the class of configurations studied here, the expectation values of arbitrary powers of the bulk edge lengths are finite.

A key point in our analysis will be the asymptotic behaviour of the Regge action for large bulk edge lengths. We will consider particular classes of configurations related to so-called Pachner moves~\cite{Pachner:1991pm}.   The Pachner moves also include spikes and configurations which we name ``spines". While spikes require a bulk vertex, spines require only a bulk edge, see the right panel of Fig.~\ref{Fig:SpikeSpine} for an illustration: Given a bulk edge, we consider the set of simplices sharing this edge, that is, the star of this edge. A spine configuration allows for this bulk edge to be arbitrarily long, while the edge lengths in the boundary of the star are fixed. Spines can appear in Lorentzian signature, whereas they cannot appear in Euclidean signature due to the Euclidean triangle inequalities. 

We will find an astonishingly simple asymptotic behaviour for the Regge action associated to the Pachner move configurations considered here. The resulting oscillatory amplitudes (for light-cone regular configurations, see the remarks below) allows for a conditional convergence of the path integral and expectation values.  This provides further evidence that Lorentzian quantum gravity models can evade the conformal factor problem~\cite{Borissova:2023izx}.

Another aspect we will reveal here, is the frequent occurrence of light-cone irregular configurations in the regime of large edge lengths. 
Light-cone irregularities generically appear for spacetimes which describe topology change~\cite{Sorkin:2019llw} and are important for the derivation of entropy from the Lorentzian path integral~\cite{Marolf:2022ybi, Dittrich:2024awu}. They are characterized by points on the manifold, with more or less than two light cones, for example trouser or yarmulke configurations. Light-cone irregularities have  also been encountered for triangulations describing cosmological evolution~\cite{Dittrich:2021gww,Asante:2021phx}, where they appeared for small bulk edge lengths only. In contrast, in this work,  we will encounter light-cone irregularities for a regime where the bulk edges can become  arbitrarily large.

Light-cone irregularities lead to imaginary terms in the action with an ambiguous  sign. Depending on this sign, the amplitudes are therefore either suppressing or enhancing.
More precisely, the Regge action has branch cuts along configurations with such light-cone irregularities. For a complete description of the Lorentzian path integral one  needs to specify whether to integrate over such configurations, and, if this is the case, on which side of the branch cuts the integration contour is placed. Clearly, in the case of infinitely long branch cuts one has to choose the side that leads to an exponential suppression of these amplitudes. Interestingly, this is the opposite choice to the one needed to obtain the correct entropy from the Lorentzian path integral, e.g., for de Sitter space~\cite{Dittrich:2024awu}.

In summary, we will find evidence that the Lorentzian (Regge) path integral is well-defined, leading to finite expectation values, and avoids the conformal factor problem of Euclidean models. However, we will also encounter a not yet well understood feature of the Lorentzian path integral, namely, light-cone irregularities in the regime of large edge lengths. This illustrates that there are still many open questions regarding the Lorentzian path integral for quantum gravity.

Our paper is structured as follows. Section~\ref{LorGeom} is devoted to the geometry of Lorentzian simplices. In Subsection~\ref{Sec:ReggeAction} we introduce the complex Regge action and the notion of light-cone (ir-)regular configurations. In Subsection~\ref{sec:gentriang} we discuss the generalized triangle inequalities which constrain the signed squared volumes of a simplex and its subsimplices. Subsequently, in Subsection~\ref{VolumeAsmp} we analyse the asymptotic scaling of the signed squared volumes in the limit of one and multiple large edges. In the second part of this work, Section~\ref{Sec:ReggeActionAsymptotics}, we study the asymptotics of the Regge action for spine and spike configurations arising in the $3-2$ and $4-1$ Pachner moves, respectively. In Section~\ref{Sec:PathIntegralAsymptotics} we consider expectation values of powers of the length variables and establish their convergence properties. We close with a discussion in Section~\ref{discussion}.

\subsection{Summary of the strategy}

The goal of quantum Regge calculus is to perform the gravitational path integral by discretizing spacetime into a simplicial manifold, and integrating over all edge-lengths $s_e$. More specifically, the path integral in Regge calculus can be written as
\begin{equation}
	Z=\int \!\! \prod_{e \subset \mathrm{bulk}}\mathrm{d}s_e \, \mu(s_e)\,\exp(i S_{\mathrm{Regge}})\,,
\end{equation}
where the boundary edges are understood to be fixed, and where $\mu(s_e)$ is a suitable measure term. 
% and where the product makes sure that all bulk edges are integrated over. 

Here we will consider triangulations for which we can decompose the path integral into a sequence of coarse graining Pachner moves \cite{Pachner:1991pm}, which in 3D are given by the $3-2$ and $4-1$ Pachner moves.  The advantage of doing so is that for both types of moves the path integral reduces to a one-dimensional integration: the $3-2$ move features only one bulk edge, whereas the $4-1$ move features four bulk edges, but also three gauge degrees of freedom, which we will gauge fix.

This strategy has been previously used in \cite{Borissova:2023izx} in a perturbative approach. Here we probe the fully non-perturbative regime. In particular we will consider the asymptotic limit of large bulk edge lengths, which often coincides with the regime where the deviations from a flat geometry are significant. It is this regime which determines the convergence of the path integral. 

Here we are dealing with Lorentzian geometries, we therefore have to consider all possible signatures for the bulk edges. The asymptotic behaviour of the Regge action will in general depend on this signature.  In particular, the signature will determine whether the action is real or complex (due to the appearance of light-cone irregular configurations) in this regime. For the cases where we have a non-vanishing imaginary term (with a non-trivial dependence on the bulk edge lengths) the path integral can be made finite by choosing the appropriate side of the branch cut for the integration contour, so that the imaginary part of the action leads to a suppression.

This leaves us with investigating the cases with real action. For those cases we will determine the finiteness of  expectation values of the form
\begin{equation}
	\braket{\lambda^n}=\int \mathrm{d}\lambda\, \mu(\lambda) \lambda^n \exp(i S_{\mathrm{Regge}})\,,
\end{equation}
(where $\lambda$ denotes the squared bulk edge length and the integral extends to $\lambda \rightarrow \infty$) via two different methods: firstly using the asymptotic form of the action and via an analytical integration, and secondly using the exact action, but using a numerical method to evaluate the path integral. Here we can allow a large class of measures $\mu(\lambda)$,  namely that $\mu(\lambda)$ is a rational function.

We will find that, with both methods, the integral is finite for expectation values of arbitrary powers of the bulk lengths. This is in stark contrast to Euclidean Regge calculus, where the $4-1$ move features an action unbounded from below. More precisely, the integrands in this case are asymptotically of the form $\lambda^n \exp(c\sqrt{\lambda})$ with a positive constant $c>0$. Thus $4-1$ moves in Euclidean Regge calculus lead  to divergent partition functions and expectation values.

 \section{Lorentzian geometry of simplices}
 \label{LorGeom}
 \subsection{ The complex Regge action }\label{Sec:ReggeAction}
 
Here we provide an overview of the Lorentzian Regge action~\cite{Sorkin:1975ah}. To this end, we will use the framework of the complex Regge action as developed in~\cite{Asante:2021phx}. A useful review of the (Euclidean, Lorentzian, and complex) simplex geometry, which includes explicit derivations of formulae for key geometric quantities, can be found in the Appendix of~\cite{Borissova:2023izx}. 
 
Regge calculus~\cite{Regge:1961px} in $d$ dimensions describes general relativity on a piecewise flat discretized manifold obtained by gluing $d$-dimensional simplices along shared $(d-1)$-dimensional subsimplices. For Lorentzian triangulations, the configuration variables of the action are the signed squared lengths  $s_e = \vec{e}\cdot \vec{e}$ of the edge vectors $\vec{e}$, with the inner product defined by the flat Minkowski metric $\eta = \text{diag}(-1,+1,\dots,+1)$. The complex Regge action~\footnote{Here we make the choice to define the complex Regge action such that it yields the Lorentzian action for a Lorentzian triangulation.} 
takes the form~\cite{Asante:2021phx,Borissova:2023izx}
 \be\label{eq:ReggeAction}
S = - \imath  \sum_{h}\sqrt{\mathbb{V}_h}\epsilon_h\,,
 \ee
 where the sum runs over bulk and boundary hinges $h$, i.e.~$(d-2)$-dimensional subsimplices, in the triangulation and $\mathbb{V}_h$ denotes their signed squared volume, which can be computed as a Caley-Menger determinant, cf.~Section~\ref{sec:gentriang}. The bulk and boundary deficit angles are defined as
 \be\label{eq:DeficitAngles}
 \epsilon_{h}^{\text{(bulk)}} = 2\pi + \sum_{\sigma \supset h} \theta_{\sigma, h}\,,\quad \epsilon_{h}^{\text{(bdry)}} = \pi   + \sum_{\sigma\supset h}\theta_{\sigma, h} \,,
 \ee
where $\theta_{\sigma,h}$ are the complex dihedral angles.

We note that the choice of the additive constant $\pi$ for the boundary deficit angle is a convention -- $\pi$ can be also replaced with $k \times\pi/2$ with $k \in \{0,1,2,3,4\}$. This selection amounts to a choice of expected boundary type, e.g., whether one expects an approximately flat boundary or rather a corner. Here we will choose $k=2$ throughout, which corresponds to a flat boundary. 
 
The complex dihedral angles $\theta_{\sigma,h}$ in a simplex $\sigma$ at a hinge $h\subset \sigma$ can be expressed as \cite{Asante:2021phx}
\ba\label{eq:DihedralAngleLog}
\theta_{\sigma,h} = -\imath \log \qty(  
\frac{ \vec{a}\cdot \vec{b} -\imath \sqrt{ (\vec{a}\cdot \vec{a}) (\vec{b}\cdot \vec{b})- (\vec{a}\cdot \vec{b})^2   }
}
{ \sqrt{\vec{a}\cdot \vec{a}}\sqrt{\vec{b}\cdot \vec{b}}}
)
\ea
where
\ba
\vec{a}\cdot \vec{b}=
\frac{d^2}{  \mathbb{V}_{h}}  \frac{\partial \mathbb{V}_{\sigma}}{\partial s_{\bar{h}}} \, ,\q\q
\vec{a}\cdot \vec{a}=  \frac{\mathbb{V}_{\rho_a} }{   \mathbb{V}_{h}  }  \, ,\q\q
\vec{b}\cdot \vec{b}=  \frac{\mathbb{V}_{\rho_b} }{   \mathbb{V}_{h}  } \, .
\ea
%\ba
%\theta_{\sigma,h} = -\imath \log\qty(\frac{
%	\frac{d^2}{  \mathbb{V}_{h}}  \frac{\partial \mathbb{V}_{\sigma}}{\partial s_{\bar{h}}}  -  \imath \,
%	\sqrt{   \frac{\mathbb{V}_{\rho_a} }{   \mathbb{V}_{h}  }\frac{\mathbb{V}_{\rho_b}}{\mathbb{V}_{h} }     -\qty( \frac{d^2}{  \mathbb{V}_{h}}  \frac{\partial \mathbb{V}_{\sigma}}{\partial s_{\bar{h}}} )^2
%	} 
%} 
%{ \sqrt{       \frac{\mathbb{V}_{\rho_a} }{   \mathbb{V}_{h}  } }   \sqrt{ \frac{\mathbb{V}_{\rho_b}}{    \mathbb{V}_{h} }} } ) \, ,
%\ee
Here $\rho_a \subset \sigma $ and $\rho_b \subset \sigma $ are the two  $(d-1)$-subsimplices sharing the hinge $h$, and $\bar{h}\subset \sigma$ is the edge opposite to $h$. 

The arguments $z$ in the square roots $\sqrt{z}$ in~\eqref{eq:DihedralAngleLog} (as well as in~\eqref{eq:ReggeAction}) can be negative, which also holds for the argument of $\log z$. We choose to use the principle branch $\arg(z)\in (-\pi,\pi)$ but need also to specify the branch cut values for the square root and the $\log$. For $r\in \mathbb{R}_+$, we define $\sqrt{-r}=\imath \sqrt{r}$ and $\log(-r)=\log(r) - \imath \pi$. In other words, for the logarithm, we adopt the branch cut value coming from the lower complex half plane, and for the square root we adopt the branch cut value coming from the upper complex half plane. 
As derived in~\cite{Asante:2021phx}, this corresponds to a choice where we complexify the  squared edge vectors as $s_e \rightarrow s_e +\imath \varepsilon$ and take the limit $\varepsilon \rightarrow 0$.

A key  advantage of using the complex dihedral angles is that they capture both Euclidean and Lorentzian angles. Both types of angles can occur in the Lorentzian Regge action: The dihedral angles $\theta_{\sigma,h}$ are constructed by projecting a $d$-simplex $\sigma$ onto the plane orthogonal to the hinge $h$, and by considering the 2D angle between the projections of the two $(d-1)$-subsimplices that meet at the hinge $h$. In the case that the hinge $h$ is timelike, the plane onto which we project is spacelike, and therefore we have a Euclidean angle. In the case that the hinge $h$ is spacelike, the plane onto which we project is timelike, resulting in a  Lorentzian angle. Note that we do not need to consider the null case, as the deficit angles are multiplied by the volume of the hinges. Therefore, null hinges do not contribute to the Regge action.

If the data $\{\vec{a}\cdot\vec{b},\vec{a}\cdot\vec{a},\vec{a}\}$ defines a Euclidean angle (i.e., these inner products can be realized as inner products between two vectors in the Euclidean plane), the complex angle~\eqref{eq:DihedralAngleLog} reproduces minus the Euclidean angle, denoted as $\theta_{\sigma,h}=-\psi^E_{\sigma,h}$. The (bulk) deficit angle therefore amounts to the usual Euclidean deficit angle $\epsilon^E_h=2\pi-\sum_{\sigma \supset h}\psi^E_{\sigma,h}$. This deficit angle is then multiplied by the square root of the negative volume squared for the (timelike) hinge. The resulting imaginary number is multiplied by $-\imath$ in (\ref{eq:ReggeAction}). This shows that the contribution of timelike hinges to the Regge action is real. 

On the other hand, we can clearly apply the same formulae to a Euclidean triangulation, i.e., a triangulation satisfying the Euclidean generalized triangle inequalities, see Section~\ref{sec:gentriang}. The Euclidean Regge action is defined as 
\ba
S^E&=& -\sum_{h}\sqrt{\mathbb{V}_h}\,\epsilon^E_h \,.
\ea
Thus, the complex Regge action evaluates to $S=\imath S^E$ for Euclidean data.

\begin{figure}[t]
	\centering
	\includegraphics[width=.9\textwidth]{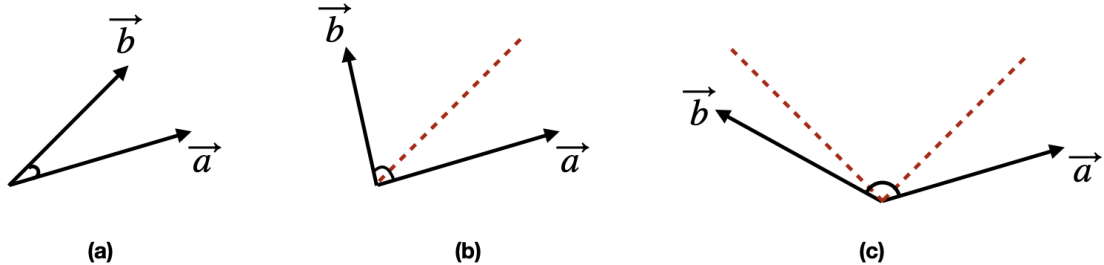}
	\hfill
	\caption{\label{Fig:angleab} 
The Lorentzian angle between $\vec{a}$ and $\vec{b}$. The dashed red lines represent the light rays. The value of the Lorentzian angle between $\vec{a}$ and $\vec{b}$ is $\left(\beta_{\sigma,h}-   \imath m_{\sigma,h} \,\pi/2\right)$, where $\beta_{\sigma,h} \in \mathbb{R}$ and $m_{\sigma,h}\in\{0,1,2\}$, depending on the number of light rays included in the (convex) wedge between $\vec{a}$ and $\vec{b}$. }
\end{figure}

Let us come back to the case of a spacelike hinge in a Lorentzian triangulation.  In this case, the data $\{\vec{a}\cdot\vec{b},\vec{a}\cdot\vec{a},\vec{a}\}$ defines a Minkowskian angle, i.e., there exist embeddings of $\vec{a},\vec{b}$ into  two-dimensional Minkowski space, which reproduce the given inner products. The complex angles (as defined in (\ref{eq:DihedralAngleLog}) will have the following structure: $\theta_{\sigma,h}=-\imath( \beta_{\sigma,h}-   \imath m_{\sigma,h} \,\pi/2)$ \cite{Sorkin:2019llw,Asante:2021phx}. Here, $\beta_{\sigma,h} \in \mathbb{R}$ and $m_{\sigma,h}\in \{0,1,2\}$ give the number of light rays included in the (convex) wedge between $\vec{a}$ and $\vec{b}$, see Fig.~\ref{Fig:angleab}.  We therefore obtain the Lorentzian (bulk) deficit angle as follows:
\ba
\epsilon^L_{\sigma,h}= 2\pi-\frac{\pi}{2}\left(\sum_{\sigma \supset h} m_{\sigma_h} \right) \,-\,\imath \sum_{\sigma \supset h} \beta_{\sigma,h} \, .
\ea
Thus, we have a purely imaginary deficit angle $\epsilon^L_{\sigma,h}$ if the sum over the $m_{\sigma,h}$ is equal to $4$, meaning that the sum of the dihedral angles includes exactly four light rays and, therefore, two light cones. We will refer to spacelike hinges that satisfy this condition as \emph{light-cone regular}. Timelike hinges are light-cone regular by definition. 

The contribution of a regular hinge to the Regge action (\ref{eq:ReggeAction}) is therefore real. On the other hand, if the number of light ray crossings in the angle associated to a given hinge $h$,
\ba
{\cal N}_h=\sum_{\sigma \supset h} m_{\sigma_h} \,,
\ea
is smaller or larger than $4$, we obtain a negative or positive imaginary contribution to the Regge action, respectively.

It is important to note that the sign of these imaginary contributions for the Lorentzian Regge action depends on the choice of conventions, as we outlined below equation~(\ref{eq:DeficitAngles}). We noted that these conventions amount to defining the Regge action by complexifying the  edge length squared as $s_e \rightarrow s_e +\imath \varepsilon$ and taking the limit $\varepsilon \rightarrow 0$. Alternatively, we can consider a complexification $s_e \rightarrow s_e -\imath \varepsilon$ and take the limit $\varepsilon \rightarrow 0$~\cite{Asante:2021phx}. This would give the opposite sign for the imaginary contributions in the Lorentzian Regge action. This shows that for all Lorentzian data which leads to light-cone irregular hinges, the Regge action (\ref{eq:ReggeAction}) has branch cuts. The Regge action on opposite sides of the branch cuts just differs in the sign of the imaginary terms coming from light-cone irregular hinges. Therefore, the choice of sign corresponds to a choice of branch cut side~\cite{Asante:2021phx}.

On the other hand, if all hinges are light-cone regular and not null, the Regge action is analytic in an open (complexified) neighbourhood around the corresponding point in the configuration space of length squares.

One might be surprised by the appearance of these imaginary terms in the Lorentzian Regge action and wonder whether an alternative definition is possible which avoids them. The imaginary terms can, however, be reproduced via analytic continuation from two different starting points: one can first apply a generalized Wick rotation and construct the Lorentzian action via analytical continuation from the Euclidean action~\cite{Jia:2021xeh,Asante:2021phx}. Second, one can start from Lorentzian data which is light-cone regular, and (using a slight deviation into complexified edge length squared to go around branch points) analytically continue to data with light-cone irregularities, see~\cite{Asante:2021phx}.

\subsection{Generalized triangle inequalities} \label{sec:gentriang}

Next, we will discuss the generalized Euclidean and Lorentzian triangle inequalities. These ensure that a simplex with given signed length squares can be embedded into flat Euclidean or Minkowskian space, respectively, and will ultimately set the integration bounds of the bulk edges in the path integral. 
The inequalities for a simplex $\sigma$ can be formulated as conditions on the signed squared volume $\mathbb{V}_{\sigma}$ for $\sigma$ and the signed squared volumes $\mathbb{V}_{\rho}$ for all its subsimplices $\rho$. The signed squared volume of a $d$-dimensional simplex $\sigma^d=(012\cdots d)$ can be computed as follows,
\ba
\mathbb{V}_{\sigma^d}&\,=\,& \frac{ (-1)^{d+1} }{ 2^{d} (d!)^2} \det  \left(
\begin{matrix}
	0 & 1 & 1 & 1 & \cdots & 1 \\
	1 & 0 & s_{01} & s_{02} & \cdots & s_{0d} \\
	1 & s_{01} & 0 & s_{12} & \cdots & s_{1d} \\
	\vdots & \vdots & \vdots & \vdots & \ddots & \vdots \\
	1 & s_{0d} & s_{1d} & s_{2d} & \cdots & 0
\end{matrix}
\right)\, ,
\ea
where $s_{ij}$ is the signed squared length of the edge between vertices $i$ and $j$. The signed squared volume determines the spacelike, null, or timelike nature of a simplex. A simplex $\sigma$ is timelike if $\mathbb{V}_{\sigma} < 0$, null if $\mathbb{V}_{\sigma} =0$, and spacelike if $\mathbb{V}_{\sigma} > 0$.

A Euclidean or spacelike (non-degenerate) simplex $\sigma$ must obey the generalized triangle inequalities,
\ba 
\mathbb{V}_{\sigma} >0,  \quad \mathbb{V}_{\rho}>0\,,
\ea 
for the simplex $\sigma$ itself and for all subsimplices $\rho$. For example, for a spacelike triangle $(012)$, the triangle inequality (for a non-degenerate triangle) requires that the squared area and the squared lengths are positive,
\ba
\mathbb{V}_{(012)}>0,\quad s_{01}>0,\quad s_{02}>0,\quad s_{12}>0 \, ,
\ea 
where the first inequality is equivalent to the well-known triangle inequality, namely that the sum of the lengths of each pair of edges has to be larger than the length of the remaining edge.

For a Lorentzian $d$-simplex $\sigma^d$ in $d$-dimensional spacetime, the subsimplices $\rho\subset \sigma^d$ can be timelike, spacelike or null. In this case, the generalized triangle inequalities~\eqref{eq:Inequalities} state that, if a subsimplex $\rho'\subset \sigma^d$ is timelike or null, then all subsimplices $\rho \subset \sigma^d$ containing this subsimplex $\rho'$, i.e., $\rho'\subset \rho$, must not be spacelike~\cite{Tate:2011rm}. In particular, a timelike subsimplex cannot be embedded in a spacelike subsimplex. Therefore, for a Lorentzian  $d$-simplex $\sigma$, we have the condition that $\mathbb{V}_{\sigma^d} < 0$ (non-degeneracy) and the following requirement: if there is a subsimplex $\rho'$ with $\mathbb{V}_{\rho'} \leq 0$, then all subsimplices $\rho$ with $\rho' \subset \rho$ need to satisfy $\mathbb{V}_{\rho} \leq 0$, i.e.,
\be\label{eq:Inequalities}
\rho \subset \sigma \,,\,\mathbb{V}_\rho \leq 0 \quad \Rightarrow \quad \forall \rho' \supset \rho: \,\mathbb{V}_{\rho'}\leq 0\,.
\ee

As an example, consider a timelike tetrahedron  $(0123)$ in three-dimensional Lorentzian spacetime with a timelike edge $(01)$ and all other edges spacelike. Then, the triangle inequalities demand
\ba 
\begin{gathered}
\mathbb{V}_{(0123)}<0,\quad \mathbb{V}_{(012)}\leq 0,\quad \mathbb{V}_{(013)}\leq 0 \, , 
\end{gathered}
\ea
as the other two triangles $(023)$ and $(123)$ can be either spacelike, null, or timelike.

The generalized triangle inequalities dictate whether we are allowed to scale one or several edges of a simplex to become large. Consider, for instance, a Euclidean triangle or more generally a Euclidean $d$-simplex. Here, we cannot scale one of the edges large while keeping the other two edges fixed, as this violates the Euclidean triangle inequality. However, such  scaling is possible for a timelike triangle (or more generally a timelike $d$-simplex). Indeed, consider the case of a timelike (non-degenerate) triangle with only spacelike edges or with only timelike edges. In this case, the Lorentzian triangle inequalities impose the opposite of the Euclidean inequality: they demand that the length of one edge is greater than the sum of the lengths of the other edges. In the case of one spacelike edge and two timelike edges (or one timelike edge and two spacelike edges), the Lorentzian triangle inequality is always satisfied.  This follows from the expression for the area square of a triangle,
\ba
\mathbb{V}_{(012)}&=&-\frac{1}{16} (s_{01}^2+s_{02}^2+s_{12}^2-2s_{01}s_{02}-2s_{01}s_{12}-2s_{02}s_{12}) \nn\\
&=&
-\frac{1}{16} (s_{01}^2+   (s_{02}-s_{12})^2-2s_{01}s_{02}-2s_{01}s_{12}) \, ,
\ea
when we consider the edge $(01)$ to be spacelike (timelike) and the other edges to be timelike (spacelike).

\subsection{Simplex geometry in the limit of large edges} \label{VolumeAsmp}

As described in the introduction, we are interested in the asymptotic behaviour of the Regge action for small simplicial complexes with boundaries, where we keep the boundary edge length fixed while scaling the bulk edge(s) to become large, if permitted by the generalized triangle inequalities.

In the following, we will focus on the geometry of a $d$-simplex $\sigma^d = (01\cdots d)$ with vertices ${0,\dots,d}$, characterized by its squared edge lengths. We will consider the asymptotic behaviour of the volume square when we scale one or several edges of this simplex to be large.

Using the asymptotic behaviour of the volumes for the $d$-simplex and its various subsimplices, we can then consider the asymptotic behaviour of the dihedral angles. To that end, we use the following formulae for the dihedral angles, which can be derived from (\ref{eq:DihedralAngleLog}), see~\cite{Dittrich:2007wm, Borissova:2023izx}:
\ba \label{SinCos} 
\sin(\theta_{\sigma, h}) \,\,=\,\, -\frac{d}{d-1}\frac{\sqrt{\mathbb{V}_h} \sqrt{\mathbb{V}_\sigma}}{\sqrt{\mathbb{V}_{\rho_a}}\sqrt{\mathbb{V}_{\rho_b}}}  \,,\q\q
\cos(\theta_{\sigma, h}) \,\,=\,\, \frac{d^2 }{\sqrt{\mathbb{V}_{\rho_a}}\sqrt{\mathbb{V}_{\rho_b}}} \pdv{\mathbb{V}_\sigma}{s_{\bar{h}}} \, .
\ea
The derivation of these relations exploits that the derivative of the squared volume with respect to a squared edge length can be expressed in terms of the squared volume  of the simplex and of its subsimplices~\cite{Dittrich:2007wm, Borissova:2023izx},
\ba
\left(\frac{\partial \mathbb{V_{\sigma}}}{\partial s_{i j}}\right)^2=\frac{1}{d^4} \mathbb{V}_{\bar{i}} \mathbb{V}_{\bar{j}}-\frac{1}{d^2(d-1)^2} \mathbb{V}_\sigma \,\mathbb{V}_{\overline{i j}}\, .
\ea
Therefore, we can compute the asymptotic behaviour of geometric quantities, such as the dihedral angles, from the asymptotic behaviour of the squared volumes of the corresponding simplex and its subsimplices.

We have already expressed the signed squared volume of a $d$-dimensional simplex $\sigma^d = (0\cdots d)$ with vertices $\{0,\dots,d\}$  via the determinant of the associated Caley-Menger matrix for the signed squared edge lengths~\cite{Sorkin:1975ah},
\ba\label{eq::VolumeCaleyMenger}
\mathbb{V}_{\sigma^d}&\,=\,& -\frac{ (-1)^{d} }{ 2^{d} (d!)^2} \det  \left(
\begin{matrix}
	0 & 1 & 1 & 1 & \cdots & 1 \\
	1 & 0 & s_{01} & s_{02} & \cdots & s_{0d} \\
	1 & s_{01} & 0 & s_{12} & \cdots & s_{1d} \\
	\vdots & \vdots & \vdots & \vdots & \ddots & \vdots \\
	1 & s_{0d} & s_{1d} & s_{2d} & \cdots & 0
\end{matrix}
\right) \equiv -\frac{ (-1)^{d} }{ 2^{d} (d!)^2} \det(C)  \,.
\ea
Making use of Laplace's expansion for the determinant of a $(d+2)\times (d+2)$ matrix $C$ and expanding around an arbitrary row $i$, we can express it as:
\be\label{eq:LaplaceDet}
\det(C) = \sum_{j=1}^{d+2} (-1)^{i+j}C_{ij} \det(\tilde{C}_{ij})\,.
\ee
Here, $C_{ij}$ denotes the $(ij)$-th entry of $C$, and $\tilde{C}_{ij}$ is the determinant of the submatrix of $C$ obtained by removing the $i$-th row and $j$-th column of $C$. Using Laplace's expansion, it is straightforward to separate the terms including large edge lengths and determine their asymptotic behaviour. 

\subsubsection{Volumes in the limit of one large edge} \label{VolumeAsmp1}

Here, we consider the situation with one large edge $(01)$.
The squared volume  $\mathbb{V}_{\sigma^d}$ of any $d$-dimensional simplex $\sigma^d=(012\cdots d)$ containing the edge $(01)$ will be a polynomial of at most quadratic order in $s_{01}$, given by $\mathbb{V}_{\sigma^d} = a\, s_{01}^2 + b\, s_{01} + c$, where $a$, $b$ and $c$ are independent of $s_{01}$. By applying Laplace's formula~\eqref{eq:LaplaceDet} repeatedly, we can determine the first coefficient as 
\be\label{eq:VolumeScalingD-2Moves}
\mathbb{V}_{(012\cdots d)} = -\frac{1}{4 d^2(d-1)^2} \mathbb{V}_{\overline{01}} \, s_{01}^2 + {\cal O}(s_{01}^{1})\,,
\ee
where $\mathbb{V}_{\overline{01}}$ is the signed squared volume of the subsimplex of $\sigma^d$ obtained by removing the vertices $(0)$ and $(1)$. 
%, and lies opposite to the edge $(01)$. 
Thus, $\mathbb{V}_{\sigma^d}$ is of order $s_{01}^2$ if $\mathbb{V}_{\overline{01}} \neq 0$. In what follows, we assume this requirement to be satisfied. 

We note that (\ref{eq:VolumeScalingD-2Moves}) relates the signature of the $d$-dimensional squared volume to the signature of $\mathbb{V}_{\overline{01}}$. That is, if $\sigma^d$ is timelike (spacelike), the subsimplex $(23\ldots d)$ needs to be spacelike (timelike). If this is not the case, the triangle inequalities do not allow us to scale the edge $(01)$ to become large. We see that scaling only one edge length to be large is only possible for timelike (or null) simplices.

For example,  with $d=2$ for a triangle $t = (012)$, we have: 
\be
\mathbb{V}_{(012)}= - \frac{1}{16} s_{01}^2 + {\cal O}(s_{01}^1)\, .
\ee
We thus see that, as discussed above, a triangle with one very large edge has to be timelike.

For a tetrahedron $\tau = (0123)$, we have:
\be\label{eq:VolumeScaling3-2Move}
\mathbb{V}_{(0123)}= - \frac{1}{144}s_{23} \, s_{01}^2 + {\cal O}(s_{01}^1)\,.
\ee
We note that this equation implies that the tetrahedron has to be timelike (or null), and $s_{23}$ has to be spacelike (or null). (If $s_{23}$ is timelike, the volume square for the tetrahedron would be positive, indicating a spacelike tetrahedron. But a spacelike tetrahedron cannot include a timelike edge.)

\subsubsection{Volumes in the limit of multiple large edges}\label{VolumeAsmpMulti}

We will consider again a $d$-simplex $(01\ldots d)$ and now scale all edges $(0i)$ with $i\in {1,\ldots,d}$ to become large.

Scaling multiple edges large requires us to specify how to do so. We could choose a multiplicative scaling $s_{0i}=\lambda\,s_{0i}^0$ or an additive scaling $s_{0i}=s_{0i}^0 \pm \lambda$ and consider the limit $\lambda \rightarrow\infty$. In the second case, the leading order  of the volumes will not retain the ``initial values"  $s_{0i}^0$, and we will indeed find simpler formulae compared to the multiplicative case. Note that the choice of additive scaling implements a form of symmetry reduction, in the sense that we consider all large squared edge lengths to have (approximately) the same modulus $\lambda$.

We start by applying an additive scaling and consider a triangle $(012)$.
Its squared volume can be written as
\ba
\mathbb{V}_{(012)} &=&
 -\frac{1}{16}s_{01}^2 - \frac{1}{16}s_{02}^2 + \frac{1}{8}s_{01}\,s_{02} + \frac{1}{8}s_{01}\,s_{12}+ \frac{1}{8}s_{02}\,s_{12}- \frac{1}{16}s_{12}^2   \, .
\ea
If $s_{01}=s_{02}=\pm\lambda$, i.e., both edges adjacent to the vertex $(0)$ are either timelike or both are spacelike, the terms quadratic in $\lambda$ cancel out. The dominant term is therefore linear in $\lambda$ and is given by
\ba
\mathbb{V}_{(012)} &=& \pm\frac{1}{4}\lambda \, s_{12} + \mathcal{O}(\lambda^0) \, .
\ea
We see that if $s_{01}=s_{02}=-\lambda$, with large $\lambda$, the triangle $(012)$ has to be timelike (or null), and therefore the edge $(12)$ has to be spacelike (or null). If $s_{01}=s_{02}=+\lambda$, the signature of the edge $(01)$ and the triangle $(012)$ need to agree.

If in turn $s_{01}=-s_{02}=\pm\lambda$, i.e., the two edges adjacent to $(0)$ have different signatures, the dominant term is quadratic in $\lambda$,
\ba
\mathbb{V}_{(012)} &=& -\frac{1}{4}\lambda^2  + \mathcal{O}(\lambda^1) \, .
\ea

%[Remark: If $s_{01}=-s_{02}=\pm\lambda$, the terms linear in $\lambda$ also cancel. However, if we change e.g. $s_{01}$ to $s_{01}a\pm\lambda$, the linear terms do not cancel anymore.]

~\\ Next, we consider the squared volume of a tetrahedron $(0123)$.
For the scaling $s_{0i} = \sigma_i\, \lambda$ with $\sigma_i=\pm 1$ and $i=1,2,3$, we can expand the volume of the tetrahedron $(0123)$ as
\ba
\mathbb{V}_{(0123)}% &=&
%\frac{1}{144} \Big((-s_{12}+s_{13}+s_{23})\sigma_{1}\sigma_{2} + (s_{12}-s_{13}+s_{23})\sigma_{1}\sigma_{3} + (s_{12}+s_{13}-s_{23})\sigma_{2}\sigma_{3}   \nn\\
% & -&\qty(s_{12}\sigma_3^2 + s_{13}\sigma_2^2 + s_{23}\sigma_1^2) \Big)\lambda^2 \nn\\
% &+& \frac{1}{144} \Big((s_{12} + s_{13} - s_{23})s_{23}\sigma_1 + (s_{12}-s_{13}+s_{23})s_{13} \sigma_2 + (-s_{12}+s_{13}+s_{23})s_{12}\sigma_3\Big)\lambda + \mathcal{O}(\lambda^0)\nn\\
& =& \frac{1}{144} \qty(\sum_{i<j, k\neq i,j} (-s_{ij} + s_{ik} + s_{jk})\sigma_i \sigma_j - s_{ij}\,\sigma_k^2)\lambda^2 
\nn\\
&&
+\frac{1}{144}  \qty(\sum_{i<j, k\neq i,j}  (-s_{ij} + s_{ik} + s_{jk})s_{ij}\,\sigma_k )\lambda 
\,\,+\,\,  \mathcal{O}(\lambda^0)\,.
\ea

If the signature of all the large edges $s_{0i}=\pm \lambda$ agrees, the terms quadratic in $\lambda$ cancel out, and we are left with
\ba\label{Vol3dA}
\mathbb{V}_{(0123)} &=& \pm \frac{1}{9} \mathbb{V}_{(123)}\,\lambda + \mathcal{O}(\lambda^0)\, .
\ea
If all large edges are timelike, the tetrahedron has to be timelike (or null), and the triangle $(123)$ has to be spacelike (or null).  In the case where all large edges are spacelike, the signature of the tetrahedron agrees with the signature of the triangle $(123)$. 
This means that in a three-dimensional Lorentzian triangulation, the triangle $(123)$ has to be timelike.

Let us now consider the case where the sign of $s_{01}=\pm \lambda$ differs from the sign of the other two edges $s_{02}=s_{03}=\mp \lambda$. The leading term is then quadratic in $\lambda$, and we have:
\ba\label{Vol3dB}
\mathbb{V}_{(0123)} &=& -\frac{1}{3^2\times 2^2} \, s_{23}\, \lambda^2 
%\pm \frac{1}{144}\large(s_{12}^2+s_{13}^2-s_{23}^2-2s_{12}s_{13}\large)\lambda 
+{\cal O}(\lambda^1)  \, .
\ea 
The tetrahedron $(0123)$ has to be timelike, and therefore $s_{23}$ has to be spacelike. All other cases with mixed signatures can be generated by renaming the vertices.\\

Let us now consider the  choice of  a multiplicative scaling of the  edges,  $s_{0i} = \lambda \, s^{0}_{0i}$ and the limit $\lambda \to \infty$. The volume of a triangle $(012)$ with squared edge lengths $s_{0i} = \lambda  \,s_{0i}^0$ where $i=1,2$ and $s_{12}$ is given by
\be
\mathbb{V}_{(012)} = -\frac{1}{16}\qty(s^0_{01}-s^0_{02})^2 \, \lambda^2 + \frac{1}{8}\qty(s^0_{01}+s^0_{02})s_{12}\,\lambda - \frac{1}{16}s_{12}^2\,.
\ee
Thus, in this case, assuming $s^0_{01}\neq s^0_{02}$ (otherwise, the multiplicative scaling is equivalent to the additive scaling for a redefined $\lambda$), the triangle will always be timelike in the limit $\lambda\rightarrow\infty$. This condition is far more restrictive than the additive scaling, where one can have spacelike triangles (if all edges involved are spacelike).  The multiplicative scaling is particularly not applicable in Euclidean signature. 

Proceeding with the squared volume of a tetrahedron with large edge lengths $s_{0i}=\lambda \, s_{0i}^0$, to leading order, it can be expressed as
\ba
\mathbb{V}_{(0123)} &=& - \frac{1}{12^2} \qty(s_{12}(s_{01}^0\! -\! s_{03}^0 )(s_{02}^0 \!-\! s_{03}^0) + s_{13}(s_{02}^0\!-\! s_{01}^0)(s_{02}^0 \!-\! s_{03}^0) + s_{23}(s_{01}^0 \!-\! s_{02}^0)(s_{01}^0\!-\! s_{03}^0) )\lambda^2\nn\\
&&+ \mathcal{O}(\lambda^1)\,.
\ea
Considering the case of three spacetime dimensions with Lorentzian signature, the tetrahedron has to be timelike. Thus, the generalized triangle inequalities are only satisfied in the large-$\lambda$ regime, if the term in the outer brackets is positive. This constitutes a rather non-trivial condition on the squared edge lengths $\{s_{12},s_{13},s_{23}\}$ and $\{s_{01},s_{02},s_{03}\}$.

In summary, we see that there are rather involved conditions to accommodate for a limit of infinite squared edge lengths with a multiplicative scaling. Therefore, we will consider only the additive scaling for the asymptotic analysis of the Regge action.

\section{Regge action asymptotics}\label{Sec:ReggeActionAsymptotics}

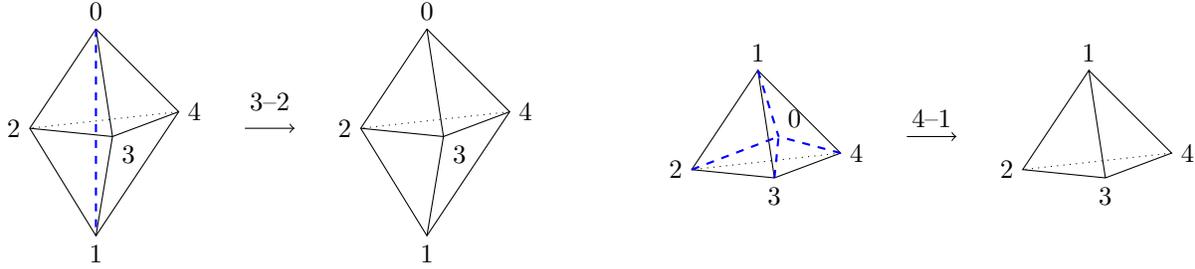
\begin{figure}[t]
	\begin{tikzpicture}[scale = 1.1]
		\draw[] (0,1.5)--(1,0.5)--(0,-1)--(-0.8,0.3)--(0,1.5);
		\draw[] (1,0.5)--(0.2,0.2)--(-0.8,0.3);
		\draw[] (0,1.5)--(0.2,0.2)--(0,-1);
		\draw[dashed,blue,thick]  (0,1.5)--(0,-1);
		\draw[dotted]  (1,0.5)--(-0.8,0.3);
		%label nodes
		\node [above] at (0,1.5) {$0$};
		\node [below] at (0,-1) {$1$};
		\node [left] at (-0.8,0.3) {$2$};
		\node [below right] at (0.2,0.2) {$3$};
		\node [right] at (1,0.5) {$4$};
		
		\draw[->] (1.8,0.3)--(2.4,0.3);
		
		\node[above] at (2.1,0.4) {3--2};
		
		\begin{scope}[xshift=4cm]
			\draw[] (0,1.5)--(1,0.5)--(0,-1)--(-0.8,0.3)--(0,1.5);
			\draw[] (1,0.5)--(0.2,0.2)--(-0.8,0.3);
			\draw[] (0,1.5)--(0.2,0.2)--(0,-1);
			\draw[dotted]  (1,0.5)--(-0.8,0.3);
			%label nodes
			\node [above] at (0,1.5) {$0$};
			\node [below] at (0,-1) {$1$};
			\node [left] at (-0.8,0.3) {$2$};
			\node [below right] at (0.2,0.2) {$3$};
			\node [right] at (1,0.5) {$4$};
		\end{scope}

	\begin{scope}[xshift=8cm,yshift=-0.5cm]
			\draw[] (0,1.5)--(1,0.5)--(0.2,0.2)--(-0.8,0.3)--(0,1.5)--(0.2,0.2);
		\draw[dotted]  (1,0.5)--(-0.8,0.3);
		\draw[dashed,blue,thick]  (0,1.5)--(0.25,0.7);
		\draw[dashed,blue,thick]  (1,0.5)--(0.25,0.7);
		\draw[dashed,blue,thick]  (-0.8,0.3)--(0.25,0.7);
		\draw[dashed,blue,thick]  (0.2,0.2)--(0.25,0.7);
		%label nodes
		\node [above right] at (0.25,0.7) {$0$};
		\node [above] at (0,1.5) {$1$};
		\node [left] at (-0.8,0.3) {$2$};
		\node [below] at (0.2,0.2) {$3$};
		\node [right] at (1,0.5) {$4$};
		
		\draw[->] (1.8,0.7)--(2.4,0.7);
		
		\node[above] at (2.1,0.7) {4--1};
		\end{scope}

		\begin{scope}[xshift=12cm,yshift=-0.5cm]
		\draw[] (0,1.5)--(1,0.5)--(0.2,0.2)--(-0.8,0.3)--(0,1.5)--(0.2,0.2);
		\draw[dotted]  (1,0.5)--(-0.8,0.3);
		%label nodes
		\node [above] at (0,1.5) {$1$};
		\node [left] at (-0.8,0.3) {$2$};
		\node [below] at (0.2,0.2) {$3$};
		\node [right] at (1,0.5) {$4$};
	\end{scope}
	\end{tikzpicture}
	\caption{\label{Fig:3DMoves}Three-dimensional Pachner moves $3-2$ and $4-1$.}
\end{figure}

Here we will consider the asymptotic behaviour of the Regge action for certain triangulations with boundaries, which contain one or several bulk edges and where we scale these bulk edges to be large.  In particular, we will consider the initial configurations of the $3-2$ Pachner move, which include one bulk edge, as well as the initial configurations of the $4-1$ Pachner move, which include four bulk edges. The $3-2$ Pachner move can lead to spine configurations, whereas the $4-1$ Pachner move can lead to spike configurations.

The initial configuration for the $3-2$ Pachner move consists of three tetrahedra sharing one bulk edge.  The boundary of this initial configuration can also serve as the boundary of two tetrahedra sharing a triangle (with a caveat discussed below). This setup represents the final configuration of the $3-2$ Pachner move, see Fig.~\ref{Fig:3DMoves}.

The initial configuration for the $4-1$ Pachner move consists of four tetrahedra sharing one bulk vertex. This initial configuration includes four bulk edges. The boundary of this initial configuration can also serve as the boundary of one tetrahedron (with a caveat discussed below), which serves as the final configuration for the $4-1$ Pachner move, see Fig.~\ref{Fig:3DMoves}. 

The final configurations of these Pachner moves, i.e., two tetrahedra sharing a triangle or a single tetrahedron, can be embedded into flat three-dimensional Minkowski space, if the generalized Lorentzian triangle inequalities hold for the tetrahedra in the final configuration. We can then construct a classical solution for the (squared) lengths of the bulk edges in the initial configurations. By using the embedding of the final configuration of a Pachner move into flat space, one can compute the lengths of the bulk edges in the initial configuration. As one uses a flat embedding, the deficit angles at all bulk edges vanish. This satisfies the three-dimensional Regge equation of motion. Consequently, the Regge action for the initial configuration evaluated on this flat solution is equal to the Regge action for the final configuration.

Note, however, that there are cases where the Lorentzian generalized triangle inequalities can be satisfied for the initial configuration of a Pachner move (with some range of edge lengths allowed for the bulk edges), but not for the final configuration. It might still be possible that the triangle inequalities are satisfied for tetrahedra with a different spacetime signature, e.g., Euclidean signature. In such cases, one can construct a solution to the equations of motion, which describes a simplicial complex of different signature. These solutions can still play a role in the path integral as saddle points in a complexified configuration space, as discussed in, for example, in ~\cite{Dittrich:2021gww,Dittrich:2023rcr,Dittrich:2024awu}.

In Lorentzian signature, there are more possibilities to allow for very large lengths of bulk edges compared to Euclidean signature. In the initial $3-2$ Pachner move configuration, the Euclidean triangle inequalities only allow for bounded bulk edge lengths (if one keeps the boundary edge lengths fixed). In contrast, we will see that with the Lorentzian triangle inequalities, one can have unbounded edge lengths and either allow for large spacelike or large timelike edges. 

The initial $4-1$ Pachner move configurations allow for unbounded bulk edge lengths in both Euclidean and  Lorentzian signature. In Lorentzian signature, we differentiate between cases where there are only spacelike or only timelike bulk edges, or a mixture of spacelike and timelike bulk edges. We will see that all cases with large spacelike edges lead to light-cone irregular configurations, and thus, in these cases, the action acquires imaginary terms.

\subsection{$3-2$ Pachner move}

The $3-2$ Pachner move starts with a configuration of three tetrahedra $(0123),(0124)$, and $(0134)$ sharing one edge $(01)$. One integrates over the squared edge length variable  $s_{01}$ and thereby ``removes the edge $(01)$", see Fig.~\ref{Fig:3DMoves}.  The classical equations of motion demand that the deficit angle associated with the edge $(01)$ vanishes, indicating that the triangulation has a flat bulk. The final configuration of this Pachner move can therefore be interpreted as two tetrahedra $(0234)$ and $(1234)$ glued along the triangle $(234)$. 

Here, we will consider the asymptotic regime for the configuration of three tetrahedra sharing an edge, particularly when this edge length is very large, which we call a ``spine" configuration. Such an asymptotic regime is not possible in Euclidean signature, where the bulk edge length is bounded by the boundary edge lengths due to the Euclidean triangle inequalities. In Lorentzian signature, however, we can have an arbitrarily large length for the bulk edge. As we keep the boundary edge lengths fixed, all triangles sharing this unbounded edge must be timelike.

\subsubsection{Two examples}\label{twoexamples}

Let us first consider two examples of such a Pachner move. These examples will illustrate the general result obtained further below. 

~\\
(A): {\bf The shared edge is timelike:}\\
We first consider the case where the shared edge is timelike.  To start, we consider one tetrahedron with a time like edge $(01)$, whose length we will let go to infinity.  We assume that the edges $(0i)$ and $(1j)$, with $i,j=2,3$, have all the same squared length, and that the edge $(23)$ is spacelike. We embed this tetrahedron into Minkowski space using the following placement of its vertices,
\ba\label{Examp1}
(0):(-t,0,0);\,\, (1):(t,0,0);\,\, (2):(0,x,b);\,\, (3): (0,x,-b) \,.
\ea
Now, if we send $t$ to infinity, we also need to adjust $x$ such that $x(t)=\sqrt{s_{02}-b^2+t^2}$, ensuring that the squared edge length $s_{02}$ (and therefore all other $s_{0i}$ and $s_{1j}$) remain constant.  This requires $s_{02}>b^2-t^2$ to obtain a real coordinate $x$. This inequality is indeed imposed by the Lorentzian triangle inequalities.

\begin{figure}[t]
	\centering
	\includegraphics[width=.48\textwidth]{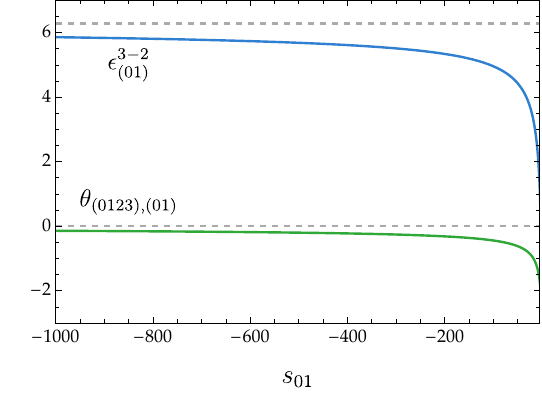}
	\hfill
	\includegraphics[width=.49\textwidth]{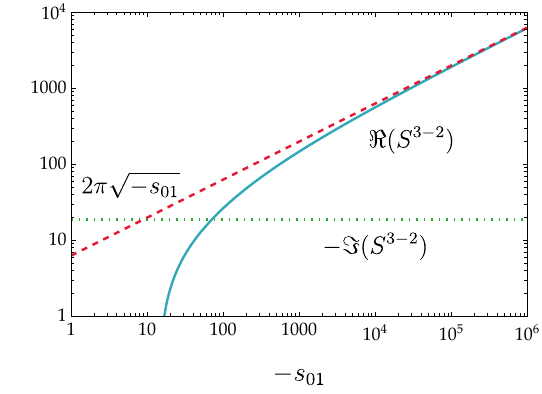}
	\caption{\label{Fig:32Example1} 
Here we consider three identical tetrahedra sharing a timelike bulk edge whose squared edge length is sent to infinity. Using the notation from equation~(\ref{Examp1}), we have $b^2=\frac{5}{4}$ and $s_{02}=1$.
In the left panel, we show the dihedral angle $\theta_{(0123),(01)}$ (green curve) and the bulk deficit angle $\epsilon^{3-2}_{(01)}$ (blue curve)  as a function of the bulk variable $s_{01}$. In the right panel, we present a logarithmic plot of the real part of the exact Regge action $S^{3-2}$ (cyan curve), its asymptotic approximation~\eqref{S32Example1} (red dashed curve), and the imaginary part of $-S^{3-2}$ (green dotted line). Note that the imaginary part results from the boundary deficit angles and can be made to vanish by redefining the boundary deficit angle, as discussed below (\ref{eq:DeficitAngles}).}
\end{figure}

The dihedral angle $\theta_{(0123),(01)}$ at the edge $(01)$ is given by (minus)  the Euclidean angle between $(0,x,b)$ and $(0,x,-b)$. It approaches $0$ in the limit as $t$ (and therefore $x$) goes to infinity. 

The dihedral angle $\theta_{(0123),(23)}$ at the edge $(23)$ can be computed as 
\ba
\theta_{(0123),(23)} &=& -\imath \log \left(\frac{b^2-s_{02}-2 t^2+2 \imath \sqrt{-t^2 \left(-b^2+s_{02}+t^2\right)}}{\sqrt{b^2-s_{02}}\sqrt{b^2-s_{02}}}\right)\nn \\
&\xrightarrow[]{t\rightarrow \infty}& -\imath \log\left(-\frac{4t^2}{\sqrt{b^2-s_{02}}\sqrt{b^2-s_{02}}}\right) = \mathcal{O}\left(\log (s_{01})\right)\,,
\ea
which scales as $\mathcal{O}(\log s_{01})=\mathcal{O}(\log t)$ for $t\to\infty$.

The remaining dihedral angles are all equal to the dihedral angle $\theta_{(0123),(02)}$ at the edge $(02)$, which can be computed as 
\ba
\theta_{(0123),(02)} &=& -\imath \log \left(\frac{-b t^2-\imath \sqrt{-s_{02} t^2 \left(-b^2+s_{02}+t^2\right)}}{\sqrt{s_{02}-b^2} \sqrt{-t^2 \left(s_{02}+t^2\right)}}\right)\nn \\
&\xrightarrow[]{t\rightarrow \infty}& -\imath \log \left(\frac{ \imath\left(b - \sqrt{s_{02}}\right)}{\sqrt{s_{02}-b^2}}\right) =\mathcal{O}(s_{01}^0)\,,
\ea
and which scales as $\mathcal{O}(s_{01}^0)$ for $t\to\infty$.

We now consider three such tetrahedra $(0123),(0124),(0134)$ which share the edge $(01)$.  With the above asymptotic behaviour of the dihedral angles and the form of the Regge action (\ref{eq:ReggeAction}), we see that the leading-order contribution to the action comes from the term  $\sqrt{s_{01}}\epsilon_{01}$. The deficit angle $\epsilon_{01}=2\pi+\sum_{i<j} \theta_{(01ij),(01)}$ with $i,j=2,3,4$ approaches $2\pi$. This leads to  
\ba\label{Case1for32}
S^{3-2}= 2\pi \sqrt{|s_{01}|}+ \mathcal{O}(\log s_{01})\,  . \label{S32Example1}
\ea
Note that the leading term  in the action is real. 

Fig.~\ref{Fig:32Example1} shows the dihedral angle and deficit angle at the edge $(01)$, for an example with $b^2=\frac{5}{4}$ and $s_{02}=1$~\footnote{Unless stated otherwise, we set $\ell_P=1$ and express geometric quantities, such as length squares and areas, in terms of Planck units.}. It also compares the exact Regge action with the asymptotic expression (\ref{Case1for32}).

~\\
(B): {\bf The shared edge is spacelike:}\\
Next, we consider a tetrahedron with a spacelike edge $(01)$ and similar symmetries as above. Here we can use embedding coordinates:
\ba\label{spacelike01}
(0):(0,-x,0);\,\,(1):(0,x,0);\,\,(2):(t,0,-b);\,\, (3):(t,0,b) \,  . \label{eq: spcelikevertices}
\ea
If we send $x$ to infinity, we need to adjust $t$ as $t(x)=\sqrt{b^2-s_{02}+x^2}$. 

The dihedral angle at $(01)$ is now a Lorentzian angle between the vectors $(t,0,-b)$ and $(t,0,b)$. This angle approaches $0$ in the limit of large $x$ (and therefore $t$).

Note that this angle is a ``thin" Lorentzian angle as it does not contain any light ray crossings.  Gluing three such tetrahedra along the edge $(01)$, the angle around $(01)$ also does not include any light ray crossings.  The edge $(01)$ is therefore light-cone irregular. (All vectors orthogonal to the edge $(01)$ are timelike; therefore, the edge represents an initial or final singularity.) As we explained in Section~\ref{Sec:ReggeAction}, light-cone irregular configurations lead to imaginary terms in the action. 

The dihedral angle $\theta_{(0123),(23)}$ at the edge $(23)$ can be computed as   
\ba
\theta_{(0123),(23)} &=& -\imath \log \left(\frac{\left(-b^2+s_{02}-2 x^2\right)-2 \imath \sqrt{-x^2 \left(b^2-s_{02}+x^2\right)}}{\sqrt{s_{02}-b^2}\sqrt{s_{02}-b^2}}\right) \nn\\
&\xrightarrow[]{x\rightarrow\infty}&    \mathcal{O}\left(\log(s_{01})\right)\,,
\ea 
and scales as $\mathcal{O}(\log s_{01})$ for $x\to\infty$.

The remaining dihedral angles are all equal to the dihedral angle $\theta_{(0123),(02)}$ at the edge $(02)$, which can be computed as
\ba 
\theta_{(0123),(02)} &=& -\imath \log \left(\frac{b x^2- \imath \sqrt{-x^2 s_{02} \left(b^2-s_{02}+x^2\right)}}{\sqrt{s_{02}-b^2}\sqrt{x^2s_{02}-x^4}}\right) \nn\\
&\xrightarrow[]{x\rightarrow\infty}&  -\imath \log \left(-\imath\frac{b+\sqrt{s_{02}}}{\sqrt{s_{02}-b^2}}\right) =  \mathcal{O}\left(s^0_{01}\right)\,,
\ea
and scales as $\mathcal{O}(s_{01}^0)$ for $x\to\infty$.

\begin{figure}[t]
	\centering
	\includegraphics[width=.48\textwidth]{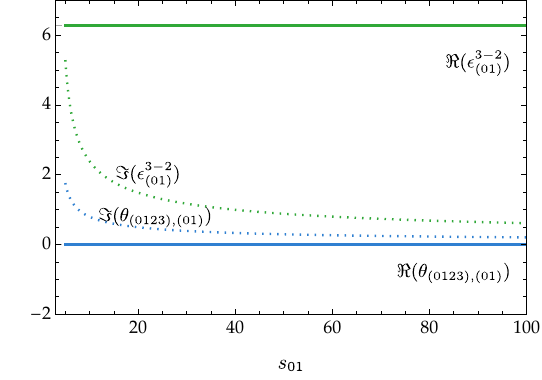}
	\hfill
	\includegraphics[width=.49\textwidth]{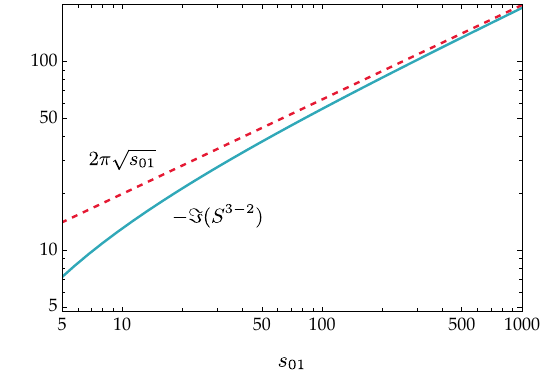}
	\caption{\label{Fig:32AnglesActionExample2}{Here, we consider three identical tetrahedra sharing a spacelike bulk edge. In the left panel, we show the  dihedral angle $\theta_{(0123),(01)}$ at the bulk edge edge $(01)$ and the bulk deficit angle $\epsilon^{3-2}_{(01)}$ at the bulk edge. In the right panel, we compare the  full Regge action~\eqref{eq:ReggeAction} with the leading-order approximation in the asymptotic regime of a large spacelike bulk edge. The boundary edge squares are fixed to $s_{0i}=s_{1i}=s_{ij}=1$, where $i,j=2,3,4$, $i<j$. The generalized inequalities~\eqref{eq:Inequalities}  are satisfied for all $s_{01}>3$.}
	}
\end{figure}

Thus, the dominant contribution in the Regge action comes again from the term $\sqrt{s_{01}}\epsilon_{01}$, and the asymptotic behaviour of the Regge action is given by 
\ba
S^{3-2}=-\imath\, 2\pi \sqrt{|s_{01}|} + \mathcal{O}(\log s_{01}) \, , \label{S32Example2}
\ea
with the leading term being imaginary. Note that with our conventions in Section~\ref{Sec:ReggeAction}, we defined the action along a specific side of the branch cuts which appear for light-cone irregular configurations. Opposite sides of the branch cuts just differ in their sign for the imaginary term.  If we consider a path integral for this configuration, we would need to integrate over the length square $s_{01}$. In this case, we can choose the integration contour along the branch cut such that the integral converges.

Fig.~\ref{Fig:32AnglesActionExample2} shows an example of such a configuration, with $s_{0i}=s_{1i}=s_{ij}=1$, for $i,j=2,3,4$. The triangle inequalities then require $s_{01}>3$. These configurations are light-cone irregular at the bulk edge for all allowed values of $s_{01}>3$.

We note that this family of configurations of three tetrahedra of type (\ref{spacelike01}) sharing an edge $(01)$ does not admit a classical solution with $s_{01}>0$.  In the case where $s_{23}=s_{24}=s_{34}$ are spacelike and $s_{0i}=s_{1i}$, for $i=2,3,4$, timelike, the triangle inequalities for the final configuration of the $3-2$ Pachner move are satisfied. But the classical solution for the bulk edge in the initial configuration of the Pachner moves demands that $s_{01}$ is timelike (with the length given by twice the height in the tetrahedron $(1234)$). 

In the case that $s_{23}=s_{24}=s_{34}$ is spacelike and $s_{0i}=s_{1i}$, for $i=2,3,4$, is spacelike, the Lorentzian triangle inequalities for the final configuration of the $3-2$ Pachner move demand that $3s_{02}<s_{23}$. In this scenario, the classical solution for the bulk edge also requires that $s_{01}$ is timelike (with the length being twice the height in the tetrahedron $(1234)$).

The family (\ref{spacelike01}) also allows configurations with $3s_{02}>s_{23}$. Thus, the Lorentzian triangle inequalities for the initial configuration of the $3-2$ Pachner move are satisfied (for an appropriate choice of the bulk length). However, for $3s_{02}>s_{23}$, the Lorentzian triangle inequalities are not satisfied for the final configuration of the Pachner move. Instead, we have a situation where the Euclidean triangle inequalities are satisfied. This means we have a ``tunneling" solution for the bulk length. The (analytically continued) Regge action evaluates to plus or minus $\imath$ times the Euclidean Regge action on such tunneling solutions. (See~\cite{Asante:2021phx} for a detailed construction of such analytically continued actions.)

\begin{figure}[t]
	\centering
	\includegraphics[width=.48\textwidth]{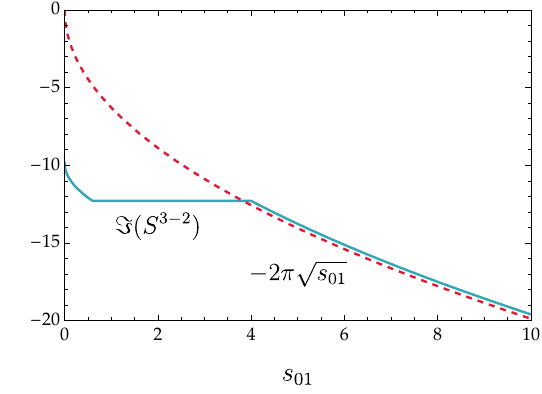}
	\hfill
	\includegraphics[width=.49\textwidth]{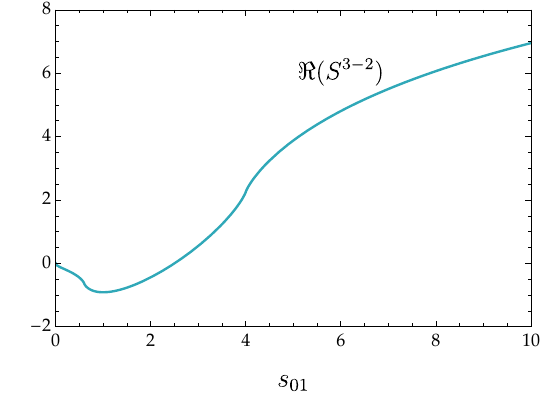}
	\caption{\label{Fig:32AnglesActionExample3} Here, we consider three tetrahedra which share a spacelike bulk edge. In the left panel, we compare the imaginary part of the full Regge action with the leading-order approximation in the asymptotic regime of a spacelike bulk edge. In the right panel, we show how the real part of the full Regge action changes with the squared bulk edge variable. The boundary edge squares are fixed as follows: $s_{02}=s_{03}=s_{12}=s_{13}=1$, $s_{04}=s_{14}=0.15$, $s_{23}=5$, and $s_{24}=s_{34}\simeq 0.20$. The generalized triangle inequalities hold for all $s_{01}>0$. The configuration is light-cone regular at the bulk edge for the range of $s_{01}$ where the imaginary part of $S^{3-2}$ is constant but becomes irregular at the bulk edge when this condition is not met.}
\end{figure}

So far, we have considered the case where the three tetrahedra sharing an edge have the same geometry. By relaxing this assumption, we can construct cases in which the configurations are light-cone regular for a certain regime of the squared edge length $s_{01}>0$. To be specific, we choose the same geometry for the tetrahedron $(0123)$ as in (\ref{eq: spcelikevertices}). To determine the geometry of the other two tetrahedra $(0124)$ and $(0134)$, we introduce a vertex $(4)$ at the coordinates $(-t', 0, 0)$, thereby constructing an embedding of the three-tetrahedral complex into flat space. We then adopt the boundary lengths from this flatly embedded configuration and vary $s_{01}$ while keeping the boundary edge lengths fixed.

Fig.~\ref{Fig:32AnglesActionExample3} shows the Regge action for an example with boundary edge lengths $s_{02} = s_{03} = s_{12} = s_{13} = 1$, $s_{04} = s_{14} = 0.15$, $s_{23} = 5$, and $s_{24} = s_{34} \simeq 0.203.$ The configurations are light-cone regular at the bulk edge for the range of $s_{01}$ where the imaginary part of the action remains constant. This constant imaginary part is caused solely by the boundary edges and can be absorbed by redefined the boundary deficit angle, see the discussion below~\eqref{eq:DeficitAngles}.

\subsubsection{General analysis}

The previous examples also capture the asymptotic behaviour in the general case. To see this, we first establish the asymptotic behaviour of the dihedral angles in a tetrahedron $(0123)$ as the squared edge length $s_{01}$ becomes large.

Using the formulae (\ref{SinCos}) for the sine and cosine of the dihedral angles in terms of volumes, and the asymptotic expressions for the volumes as discussed in Section~\ref{VolumeAsmp}, we find:

\begin{itemize}\label{eq:DihedralAngle32MoveAsymptotics}
\item {\it Dihedral angle at the (bulk) edge $(01)$:}
\ba
\sin(\theta_{(0123),(01)}) &=& -\frac{3}{2} \frac{\sqrt{s_{01}}\sqrt{-\frac{1}{144}s_{23}s_{01}^2}}{\sqrt{-\frac{1}{16}s_{01}^2}\sqrt{-\frac{1}{16}s_{01}^2}} +\mathcal{O}(s_{01}^{-3/2}) \,=\, \mathcal{O}\qty(s_{01}^{-1/2}) \to \,0\,, \quad s_{01}\to \pm \infty\,.\nn\\
\ea
On the other hand, if we consider the cosine of the same angle, we
 obtain
\ba
\cos(\theta_{(0123),(01)}) 
&=& 
3^2 \frac{-\frac{1}{144}s_{01}^2}{\sqrt{-\frac{1}{16}s_{01}^2}\sqrt{-\frac{1}{16}s_{01}^2}}
+ \mathcal{O}(s_{01}^{-1}) \,=\, 1+ \mathcal{O}(s_{01}^{-1}) \, .
\ea
Thus the dihedral angle at the edge $(01)$ goes to zero in the limit of a large squared edge length $s_{01}\rightarrow\pm \infty$. This behaviour is illustrated in Fig.~\ref{Fig:32Example1} (left) for a specific choice of edge lengths compatible with the generalized triangle inequalities~\eqref{eq:Inequalities}.

\item {\it Dihedral angles at the (boundary) edges $(0i)$ and $(1i)$, $i=2,3$:}

Consider for example
\ba
 \sin(\theta_{(0123),(02)})&  =& -\frac{3}{2} \frac{\sqrt{s_{02}}\sqrt{-\frac{1}{144}s_{23}s_{01}^2}}{\sqrt{\mathbb{V}_{(023)}}\sqrt{-\frac{1}{16}s_{01}^2}} +\mathcal{O}(s_{01}^{-1})  \,=\, \mathcal{O}\qty(s_{01}^0)\,, \quad s_{01}\to \pm \infty\,.
 \ea

In the Regge action, this boundary dihedral angle is multiplied by the boundary edge length. This leads to a $\mathcal{O}(s_{01}^0)$ term, which is subleading compared to the term coming from the bulk edge.

\item {\it Dihedral angle at the (boundary) edge $(23)$:}
 \ba
  \sin(\theta_{(0123),(23)})&=& -\frac{3}{2} \frac{\sqrt{s_{23}}\sqrt{-\frac{1}{144}s_{23}s_{01}^2}}{\sqrt{\mathbb{V}_{(023)}}\sqrt{\mathbb{V}_{(123)}}} +\mathcal{O}\qty(s_{01}^0)\, =\, \mathcal{O}\qty(s_{01}^1)\,, \quad s_{01}\to \pm \infty \,.
 \ea
The dihedral angle $\theta_{(0123),(23)}$ therefore grows as $\mathcal{O}(\log(s_{01}))$. (Remember that $\arcsin(x) = -\imath \log(\imath x + \sqrt{1-x^2})$.) This still leads to a subleading term in the Regge action, compared to the term coming from the bulk edge.
\end{itemize}

We can now proceed to determine the asymptotic behaviour for the deficit angles. Remembering that these are defined as $ \epsilon_{h}^{\text{(bulk)}} = 2\pi + \sum_{\sigma \supset h} \theta_{\sigma, h}$ and $\epsilon_{h}^{\text{(bdry)}} = \pi   + \sum_{\sigma\supset h}\theta_{\sigma, h}$, we obtain the following:

\begin{itemize}
	\item {\it Deficit angle at the bulk edge $(01)$:}
	\be\label{eq:DeficitAngle32MoveAsymptotics1}
	\epsilon_{(01)} = 2\pi + \theta_{(0123),(01)} +  \theta_{(0124),(01)}+ \theta_{(0134),(01)} \to \, 2\pi\,,\quad  s_{01}\to \pm \infty\,.
	\ee
Thus, as a consequence of the three-dimensional dihedral angles approaching zero asymptotically, the bulk deficit angle approaches $2\pi$. This is illustrated in Fig.~\ref{Fig:32Example1} (left) for a specific choice of edge lengths. 
 
	\item {\it Deficit angles at the boundary edges $(0i)$ and $(1i)$, $i=2,3,4$:}
	\be\label{eq:DeficitAngle32MoveAsymptotics2}
	\epsilon_{(02)} = \pi + \theta_{(0123),(02)} +  \theta_{(0124),(02)}= \mathcal{O}\qty(s_{01}^0) \,,\quad  s_{01}\to \pm \infty\,.
	\ee
	\item {\it Deficit angle at the boundary edges $(ij)$, $i,j=2,3,4$, $i<j$:}
	\be\label{eq:DeficitAngle32MoveAsymptotics3}
	\epsilon_{(23)} = \pi + \theta_{(0123),(23)}= \mathcal{O}\qty(\log(s_{01})) \,,\quad  s_{01}\to \pm \infty\,.
	\ee
\end{itemize}

Thus, with the Regge action given by $
S= \sum_{h}\sqrt{\mathbb{V}_h}\epsilon_h$, we conclude  
\ba\label{eq:ReggeAction32MoveAsymptotics}
 S^{3-2} &=&-\imath \, \sqrt{s_{01}}\,\epsilon_{(01) } + \mathcal{O}(\log s_{01}) \,=\,-\imath \, 2\pi \sqrt{s_{01}} + \,\, \mathcal{O}(\log s_{01}) \,, \quad s_{01}\to \pm \infty\,.
\ea

We thus confirm the behaviour found in the two examples of Section~\ref{twoexamples}. For the case of a shared timelike edge, the leading-order term in the Regge action is real. For the case of a shared spacelike edge, the leading-order term is imaginary. This reflects that the spacelike bulk edge is light-cone irregular, as the angle around it includes zero light rays.

\subsection{Generalization to $N$ tetrahedra sharing an edge}\label{Gen32}

We can easily generalize the considerations for the $3-2$ Pachner move configuration to a configuration of $N$ tetrahedra sharing an edge. In fact, the asymptotic behaviour for the Regge action is given by
\ba\label{eqGen32}
 S^{N\, \text{tetra}} &=&-\imath\,\sqrt{s_{01}}\epsilon_{(01) } + \mathcal{O}(\log s_{01}) \,=\,-\imath \, 2\pi \sqrt{s_{01}} + \,\, \mathcal{O}(\log s_{01}) \,, \quad s_{01}\to \pm \infty\,,
\ea
and thus remains the same as for the  $3-2$ Pachner move configuration, i.e., independent of $N$.

We note that this asymptotic form of the action might have peculiar consequences for the asymptotic form of the  path integral amplitudes for the Lorentzian Ponzano-Regge model. To this end, let us first note that one can define a phase space for Regge calculus~\cite{Dittrich:2009fb,Dittrich:2011ke}. In the $(2+1)$-dimensional theory, length variables are conjugated to boundary deficit angles. For a timelike edge, the boundary deficit angle is compact, and one therefore expects the length operator to have a discrete spectrum, which for large lengths is equidistant. For a spacelike edge, the boundary deficit angle is non-compact, and one expects the length operator to have a continuous spectrum. These expectations are indeed satisfied for the Lorentzian Ponzano-Regge model~\cite{Davids:2000kz,Freidel:2002hx,Freidel:2000uq}, where the spectrum for the timelike length goes like $T\sim j$ for large $j$ and $j\in \mathbb{N}$. Thus, ignoring (constant) boundary terms and measure terms, the quantum-mechanical amplitudes behave as $\sim\exp(\imath\, 2\pi j) = 1$. Let us also remark that exact one-loop measure terms can be derived~\cite{Dittrich:2011vz,Borissova:2023izx}, and they tend to suppress amplitudes for large edge lengths, as discussed in~Section~\ref{Sec:PathIntegralAsymptotics}.

\subsection{$4-1$ Pachner move}

The $4-1$ Pachner move starts with a configuration of four tetrahedra $(0123),(0124)$, $(0134)$ and $(0234)$ sharing one vertex $(0)$. One integrates over the edge square variables  $s_{01},s_{02},s_{03}$ and $s_{04}$ and in this way ``removes the bulk edges", see Fig.~\ref{Fig:3DMoves}.  The classical equations of motion demand that the deficit angles associated with the edges $(0i)$, where $i=1,\ldots,4$ vanish, i.e., indicating a flat bulk triangulation. The final configuration of this Pachner move can therefore be interpreted as one tetrahedron $(1234)$. 

We will now consider the asymptotic regime for the configuration of four tetrahedra sharing one vertex, with the bulk edge lengths being very large.  We need to take into account all possible signatures of the bulk edges, that is, the homogeneous case where all four bulk edges are spacelike or timelike, as well as the inhomogeneous case where a subset of the bulk edges is spacelike, while the others are timelike.

As mentioned previously, the equations of motion for the bulk edge lengths demand a flat configuration. Such solutions can be easily constructed: If the generalized Lorentzian triangle inequalities for the tetrahedron $(1234)$ are satisfied, we can embed this tetrahedron into Minkowski space. Furthermore, by embedding a vertex $(0)$ inside this tetrahedron, we can construct a three-parameter family of solutions. (If the boundary data satisfies the Euclidean triangle inequalities, we can construct  Euclidean solutions which define a family of saddle points for the complexified Regge action~\cite{Asante:2021phx}.)  This three-parameter family of solutions constitutes one gauge orbit with flat configurations.  The gauge symmetry in question can be identified as a remnant of diffeomorphism symmetry~\cite{Dittrich:2008pw,Bahr:2009ku}. The Regge action evaluates to the same value on this gauge orbit and coincides with the Regge action of the tetrahedron  $(1234)$. 

The notion of gauge orbits extends to curved configurations: we define configurations with the same value of the Regge action as belonging to the same gauge orbit. Note that, as we have four bulk variables and consider one condition, namely a constant Regge action, the gauge orbits are generically three-dimensional. 

For the path integral, we are supposed to integrate over all four bulk lengths. But since we have a three-dimensional gauge symmetry~\footnote{From the description of the gauge orbit for the flat solution above, one would expect that this gauge orbit is compact. One can, however, introduce the orientation of the top-dimensional simplices as a further summation variable. This allows for non-compact gauge orbits~\cite{Freidel:2002dw,Christodoulou:2012af}.}, we can use a gauge fixing and reduce the path integral to a one-dimensional integral. 

Here we will consider a gauge fixing for the asymptotic regime, in which we set the modulus for the bulk edge lengths to be equal, i.e., $|s_{0i}|=\lambda$ for $i=1,\ldots,4$. This choice of gauge fixing corresponds to the additive scaling discussed in Section~\ref{VolumeAsmpMulti}. Using this additive scaling as gauge fixing, we have to assume that the gauge conditions $|s_{0i}|=\lambda$ define a good gauge fixing for large $\lambda$. We will find that the action is linear in $\sqrt{\lambda}$ in the asymptotic regime, which is consistent with this assumption.

We will proceed by considering first the case of homogeneous signature for all the bulk edges.

~\\
{\bf Case $s_{0i}=\pm \lambda$ (with the same sign for all $i=1,\ldots,4$):}\\\noindent
To compute the Regge action, we  first determine the asymptotic limit of the dihedral angles.  For the dihedral angle at an  edge $(0i)$, we compute
\ba\label{eq66}
\sin(\theta_{(0123),(01)}) \,=\, -\frac{3}{2} \frac{\sqrt{s_{01}}\sqrt{\mathbb{V}_{0123}}}{\sqrt{\mathbb{V}_{012}}\sqrt{\mathbb{V}_{013}}} 
&=& - 2\frac{\sqrt{\pm \lambda}\sqrt{\pm\mathbb{V}_{123}\lambda}}{\sqrt{\pm s_{12}\lambda}\sqrt{\pm s_{13}\lambda}} 
+\mathcal{O}(\lambda^{-1})\nn\\
&=&  \sin(\theta_{(123),(1)}) +\mathcal{O}(\lambda^{-1}) \,  ,
\ea
and similarly,
\ba\label{eq67}
\cos(\theta_{(0123),(01)}) &=& \cos(\theta_{(123),(1)}) +\mathcal{O}(\lambda^{-1}) \, .
\ea

For the dihedral angle at an  edge $(ij)$, where $i,j=1,2,3$, we compute
\ba
\sin(\theta_{(0123),(12)}) \, =\,
-\frac{3}{2} \frac{\sqrt{s_{12}}\sqrt{\mathbb{V}_{0123}}}{\sqrt{\mathbb{V}_{012}}\sqrt{\mathbb{V}_{123}}}
&=& - \frac{\sqrt{s_{12}}\sqrt{\pm\mathbb{V}_{123}\lambda}}
{\sqrt{\pm s_{12}\lambda}\sqrt{\mathbb{V}_{123}}}\,+\,\mathcal{O}(\lambda^{-1}) \nn\\
&=&  -1 \,+\,\mathcal{O}(\lambda^{-1}) \, .
\ea
Thus, we find that the angle $\theta_{(0123),(12)} =-\pi/2 +\mathcal{O}(\lambda^{-1/2})$.
(Note that we have $\mathcal{O}(\lambda^{-1/2})$  because $-1$ is a special expansion point for $\arcsin$.)

~\\
For the asymptotic limit of the action, we need to consider the deficit angles at the bulk edges $(0i)$, $i=1,2,3,4$. For the edge $(01)$, we obtain
\ba\label{eq:deficit41}
	   \epsilon_{(01)}^{4-1} &  =&2\pi + \theta_{(0123),(01)} +   \theta_{(0124),(01)}+ \theta_{(0134),(01)} \nn \\
	   &=& 2\pi + \theta_{(123),(1)} +\theta_{(124),(1)} +\theta_{(134),(1)} + \mathcal{O}(\lambda^{-1}) \, ,
\ea
Similar calculations can be done for the other edges $(0i)$, where $i=2,3,4$. Thus, the deficit angles at the four bulk edges will include all two-dimensional angles which appear in the four triangles forming the boundary of the tetrahedron $(1234)$.  In the Regge action, all of these deficit angles are multiplied by the same coefficient, namely $-\imath\,\sqrt{\pm \lambda}$. By summing over the angles in the four triangles of the tetrahedron, we obtain a term of $-\imath\,\sqrt{\pm \lambda}(8\pi - 4\pi)$, as the angles in a triangle sum up to $-\pi$.

As for the boundary deficit angles, where $i,j,k,l=1,2,3,4$ are pairwise different, we obtain
		\ba
		\epsilon_{(ij)}^{4-1} &  =& \pi + \theta_{(0ijk),(ij)} +   \theta_{(0ijl),(ij)}  \nn\\
  &=&  \mathcal{O}(\lambda^{-1/2}) \, .
		\ea
These boundary deficit angles will not contribute to leading order in the Regge action.

\begin{figure}[t]
	\centering
	\includegraphics[width=.48\textwidth]{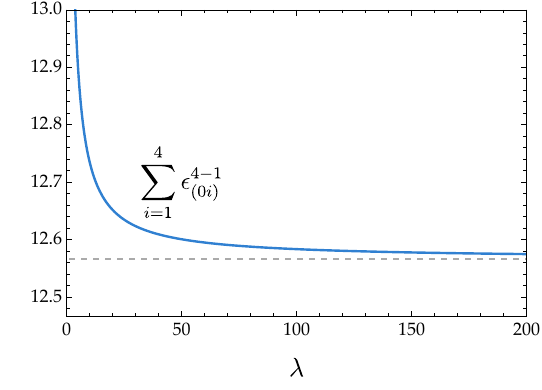}
	\hfill
	\includegraphics[width=.49\textwidth]{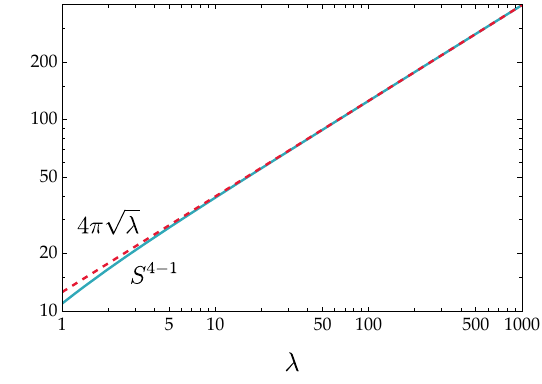}
	\caption{\label{Fig:41moveexple1}  
Here we consider a $4-1$ Pachner move configuration with timelike bulk edges with length square $s_{0i}=-\lambda$. All boundary edges are spacelike and have length square $s_{ij}=1$.
 Left: The sum  over the bulk deficit angles $\epsilon^{4-1}_{(0i)}$ (blue curve) as a function of $\lambda$. Right: Log-log plot of the real part of the exact Regge action $S^{4-1}$ (cyan curve)  and the asymptotic approximation~\eqref{eq:ReggeAction41MoveAsymptotics} (red dashed curve) as functions of $\lambda$.}
\end{figure}

Hence, overall we obtain for the asymptotic limit of the Regge action:
\ba\label{eq:ReggeAction41MoveAsymptotics}
S^{4-1} &=&- 4\pi \imath  \, \sqrt{\pm \lambda} \,+\, \mathcal{O}(\lambda^0) \, .
\ea
We see that for timelike bulk edges, the leading term in the action is real. Indeed, timelike edges are always light-cone regular. Fig.~\ref{Fig:41moveexple1} shows the sum over bulk deficit angles and the Regge action for an explicit example with timelike bulk edges.

In the case where all bulk edges are spacelike (and the triangle inequalities can be satisfied), the asymptotic regime is light-cone irregular. Indeed, if all bulk edges are spacelike and large, all boundary triangles need to be timelike. (See the remark below~\eqref{Vol3dA}.) Therefore, the boundary triangulation defines a two-dimensional Lorentzian triangulation.  A bulk edge $(0i)$ is asymptotically irregular if and only if the vertex $(i)$ is irregular with respect to the two-dimensional (Lorentzian) geometry defined on the boundary of the tetrahedron $(1234)$.  As topological two-spheres do not admit regular Lorentzian geometries, some or all of these vertices have to be irregular. 

Coming back to the case of timelike bulk edges, we can make a similar remark as in Section~\ref{Gen32} regarding the path integral amplitudes in the Lorentzian Ponzano Regge model. With a discrete spectrum $\sqrt{|\lambda|}\sim j$, with $j \in \mathbb{N}/2$, the amplitudes become $\exp(\imath S^{4-1})\sim 1$.

~\\\noindent
Next, we consider the cases where the bulk edges have different signatures.

~\\\noindent
{\bf Case $s_{01}=\pm \lambda$  and $s_{0j}=\mp \lambda$ for $j=2,3,4$:}\\\noindent
Here we have tetrahedra of two different types. First, there is the tetrahedron $(0234)$ with all edges $(0i)$, where $i=2,3,4$, having the same squared edge length $s_{0i}=\mp\lambda$.  And second, there are the three tetrahedra $(01ij)$ with $i<j$ and $i,j=2,3,4$, where we have $s_{01}=\pm \lambda$ and $s_{0i}=\mp \lambda$. For the latter, we need to distinguish between the edges $(01)$ and $(0i)$, as well as between $\{(1i),(1j)\}$ and $(ij)$.

\begin{figure}[t]
	\centering
	\includegraphics[width=.48\textwidth]{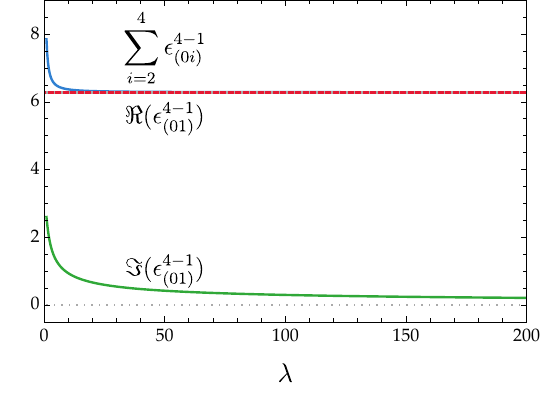}
	\hfill
	\includegraphics[width=.49\textwidth]{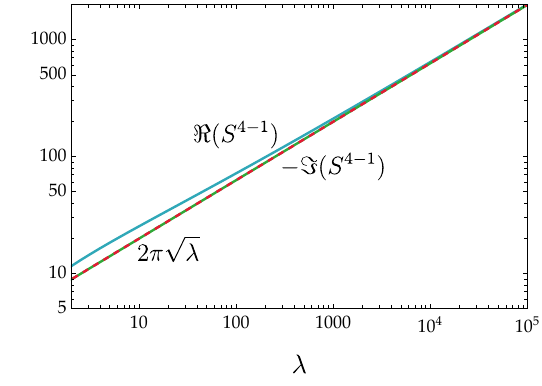}
	\caption{\label{Fig:41moveexple2}  Here we consider a $4-1$ Pachner move configuration with boundary edges $s_{ij}=1$, one spacelike bulk edge $s_{01}=\lambda$, and three timelike bulk edges $s_{02}=s_{03}=s_{04}=-\lambda$. In the left panel, we show the sum of the bulk deficit angles $\sum_{i=2}^{4}\epsilon_{(0i)}^{4-1}$ (blue curve), as well as the real (red line) and imaginary (green curve) part of the deficit angle $\epsilon_{(01)}^{4-1}$ as functions of $\lambda$, along with their asymptotic approximation (\ref{eqn:Asymptotic4-2deficit}) (gray dotted lines). In the right panel we show a log-log plot of the real part of the exact Regge action $S^{4-1}$ (teal curve), the imaginary part of $-S^{4-1}$ (green line), and the asymptotic approximation ~\eqref{eq:ReggeAction41MoveAsymptotics2} (red dashed curve) as functions of $\lambda$.}
\end{figure}

We start by considering the dihedral angle at the edge $(01)$ in the tetrahedra $(01ij)$, for which we obtain
\ba
\sin(\theta_{(01ij),(01)}) \,=\,  -\frac{3}{2} \frac{\sqrt{s_{01}}\sqrt{\mathbb{V}_{01ij}}}{\sqrt{\mathbb{V}_{01i}}\sqrt{\mathbb{V}_{01j}}} \,=\, -\frac{3}{2} \frac{\sqrt{\pm \lambda}\sqrt{-\frac{1}{36}s_{ij}\lambda^2}}{\sqrt{-\frac{1}{4}\lambda^2}\sqrt{-\frac{1}{4}\lambda^2}} + \mathcal{O}(\lambda^{-3/2}) \,=\, \mathcal{O}(\lambda^{-1/2}) \,.
\ea
Similarly, we find $\cos(\theta_{(0123),(01)}) = 1 +\mathcal{O}(\lambda^{-1})$. Therefore, we have
\ba\label{eq74}
\theta_{(01ij),(01)}&=&0+\mathcal{O}(\lambda^{-1/2}) 
\ea
in the limit $\lambda\to +\infty$.

Now let us consider the dihedral angles at the edges $(0j)$ with $j = 2,3,4$ in the tetrahedra $(01ij)$ with $i=2,3,4$.  For instance,
\ba
\sin(\theta_{(0123),(02)}) \,=\,  -\frac{3}{2} \frac{\sqrt{s_{02}}\sqrt{\mathbb{V}_{0123}}}{\sqrt{\mathbb{V}_{012}}\sqrt{\mathbb{V}_{023}}} 
\,=\,
 -\frac{3}{2} \frac{\sqrt{\mp\lambda}\sqrt{-\frac{1}{36}s_{23}\lambda^2}}{\sqrt{-\frac{1}{4}\lambda^2}\sqrt{\mp  \frac{1}{4}\lambda s_{23}}}
\,=\,-1+\mathcal{O}(\lambda^{-1}) \, ,
\ea
similarly, we find $\cos(\theta_{(0123),(02)}) = \mathcal{O}(\lambda^{-1/2})$. We thus have
\ba\label{eq76}
\theta_{(01ij),(0j)}&=&-\pi/2+\mathcal{O}(\lambda^{-1/2}) 
\ea
in the limit $\lambda\to +\infty$.

For the dihedral angles at the edges $(0j)$ with $j=2,3,4$ in the tetrahedron $(0234)$, we can use equations~(\ref{eq66},\ref{eq67}) and obtain
 \ba
\theta_{(0234),(0j)}&=& \theta_{(234),(j)} +\mathcal{O}(\lambda^{-1})\,.
 \ea

One can also find that the dihedral angles at the edges $(1i)$ or $(ij)$ grow at most with $\log \lambda$, and thus do not contribute to the leading term in the Regge action.  

For the deficit angles at the bulk edges we obtain
\ba\label{eqn:Asymptotic4-2deficit}
\epsilon_{(01)}^{4-1} &  =&2\pi + \theta_{(0123),(01)} +   \theta_{(0124),(01)}+ \theta_{(0134),(01)} \;\,=\, 2\pi  +\mathcal{O}(\lambda^{-1/2}) \, ,\nn\\
\epsilon_{(0i)}^{4-1} &  =&2\pi + \theta_{(01ij),(0i)} +   \theta_{(01ik),(0i)}+ \theta_{(0234),(0i)} \q\,=\, 2\pi  - \frac{\pi}{2}  - \frac{\pi}{2}+  \theta_{(234),(i)} + \mathcal{O}(\lambda^{-1/2}) \, ,\;\;
\ea
where $2\leq i,j,k\leq 4$ and $i,j,k$ are pairwise different. The three deficit angles $\epsilon_{(0i)}^{4-1}$ with $i=2,3,4$ include the three angles $\theta_{(234),(i)}$ in the triangle $(234)$, which sum up to $-\pi$. 

We therefore have for the Regge action
\ba\label{eq:ReggeAction41MoveAsymptotics2}
S^{4-1} &=& - \imath \sum_{i}\sqrt{s_{0i}}\epsilon_{0i}^{4-1} + \mathcal{O}(\log\lambda) \nn\\
& =& - 2\pi \imath \,\qty(\sqrt{\pm \lambda} + \sqrt{\mp \lambda}) +  \mathcal{O}(\log\lambda)\nn\\
&=& - 2 \pi \imath \,\qty(1+\imath)\sqrt{\lambda}+  \mathcal{O}(\log\lambda) 
\,.
\ea
We see that in the asymptotic limit we obtain light-cone irregular configurations. If $(01)$ is spacelike, this edge is light-cone irregular. If, instead, the three edges $(0i)$  with $i=2,3,4$ are spacelike, then at least one of these edges is light-cone irregular. 

Fig.~\ref{Fig:41moveexple2} shows the exact Regge action and its asymptotic approximation for the case of one spacelike and three timelike bulk edges.

~\\	\noindent
{\bf Case $s_{01}=s_{02}=\pm \lambda$  and $s_{03}=s_{04}=\mp \lambda$:}\\\noindent
Here all tetrahedra have either two large spacelike edges and one large timelike edge, or two large timelike edges and one large spacelike edge. The dihedral angles for the edges $(0i)$ are given in eq.~(\ref{eq74}) (if $(0i)$ differs in signature from the other two edges $(0j)$ and $(0k)$) and in eq.~(\ref{eq76}) (if $(0i)$ agrees in signature with one of the other two edges $(0j)$ or $(0k)$). They approximate the values $0$ and $-\pi/2$ for $\lambda \rightarrow \infty$, respectively.

\begin{figure}[t]
	\centering
	\includegraphics[width=.48\textwidth]{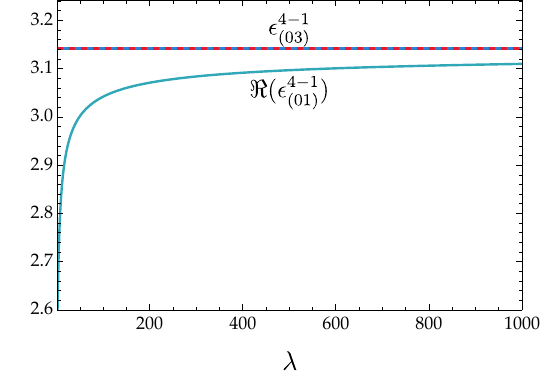}
	\hfill
	\includegraphics[width=.49\textwidth]{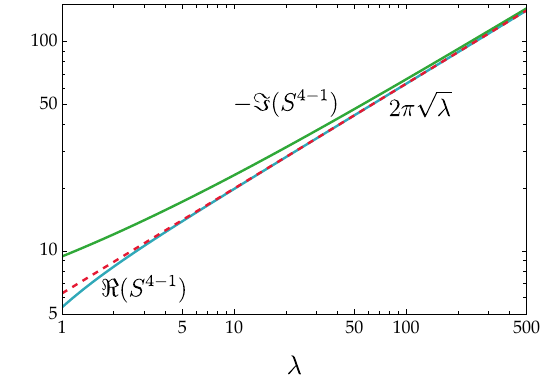}
	\caption{\label{Fig:41moveexple3}  Here we consider a $4-1$ Pachner move configurations with boundary edges $s_{ij}=1$ and with two spacelike bulk edges $s_{01}=s_{02}=\lambda$ and two timelike bulk edges $s_{03}=s_{04}=-\lambda$. In the left panel, we show the bulk deficit angle $\epsilon_{(03)}^{4-1}$ (blue curve) and the real part of the deficit angle $\epsilon_{(01)}^{4-1}$ (teal curve) as a function of the variable $\lambda$. The asymptotic approximation $\pi$ in (\ref{eqn:Asymptotic4-2deficit3}) to both deficit angles is indicated with a red dashed line. In the right panel, we show a log-log plot of the real part of the exact Regge action $S^{4-1}$ (teal curve), the imaginary part of $-S^{4-1}$ (green curve), and the analytical approximation~\eqref{eq:ReggeAction41MoveAsymptotics3} (red dashed curve) as a function of $\lambda$.}
\end{figure}

Thus we obtain for the deficit angles at the edges $(0i)$, $i=1,2,3,4$
\ba \label{eqn:Asymptotic4-2deficit3}
\epsilon_{(0i)}^{4-1} &  =&2\pi + \theta_{(0ijk),(0i)} +   \theta_{(0ijl),(0i)}+ \theta_{(0ilk),(0i)} 
%\,=\, 2\pi  - \frac{\pi}{2}  - \frac{\pi}{2} + \mathcal{O}(\lambda^{-1/2})
\,=\, \pi  + \mathcal{O}(\lambda^{-1/2})\, .\q
\ea

The dihedral angles at the edges $(ij)$ with $j>i>0$ are at most of order $\mathcal{O}(\log \lambda)$, and thus do not contribute to the leading order of the Regge action. 

We obtain for the Regge action
\ba\label{eq:ReggeAction41MoveAsymptotics3}
S^{4-1} &=& - \imath \sum_{i}\sqrt{s_{0i}}\epsilon_{0i}^{4-1} + \mathcal{O}(\log\lambda) \nn\\
& =& - 2\pi \imath \, \qty(\sqrt{\pm \lambda} + \sqrt{\mp \lambda}) +  \mathcal{O}(\log\lambda)\nn\\
&=& - 2 \pi \imath \, \qty(1+\imath)\sqrt{\lambda}+  \mathcal{O}(\log\lambda)
\,\,\,.
\ea
Here we conclude that, in the asymptotic limit,  all spacelike bulk edges are light-cone irregular.

Fig.~\ref{Fig:41moveexple3} illustrates the accuracy of the asymptotic approximation for a Pachner move with boundary length squares $s_{ij}=1$.

\subsection{Comparison with linearized action}

In the previous subsections, we have derived approximations to the Regge action in the asymptotic regime of large bulk edges for Pachner move configurations. We found that configurations, where one or more of the bulk edges are spacelike, are light-cone irregular in the asymptotic regime. In this section we wish to restrict to cases with a real Regge action, and therefore consider only timelike bulk edges. In this case, we derived for the $3-2$ Pachner move configuration the asymptotics
\ba\label{time32}
S^{3-2} &=& 2 \pi \sqrt{\lambda} + \mathcal{O}(\log(s_{01})) \,,
\ea
where $s_{01}=-\lambda$ is the bulk squared edge length. For the $4-1$ Pachner move configuration, we found
\ba\label{time41}
S^{4-1} &=&  4\pi  \sqrt{\lambda} + \mathcal{O}(\lambda^0) \,,
\ea
where $s_{0i}=-\lambda$, with $i=1,\ldots,4$, are the bulk squared edge lengths.

Let us compare these asymptotic expressions to the linearized action around a classical (and therefore Minkowski flat) solution. Such a linearized action has been derived in~\cite{Borissova:2023izx}. For the $3-2$ Pachner move, the linearized action\footnote{We omit here the part of the action which is constant in the bulk fluctuations.} is given by
\ba
S^{3-2}_{\text{qu}}  &=&  \frac{\tilde\lambda^2}{24} 
\frac{\sqrt{\abs{\mathbb{V}_{(0234)}}} \sqrt{\abs{\mathbb{V}_{(1234)}}} }
{
\sqrt{\abs{\mathbb{V}_{(0123)}}} \sqrt{\abs{\mathbb{V}_{(0124)}}} \sqrt{\abs{\mathbb{V}_{(0134)}}} 
}\,,
\ea
where $\tilde \lambda =\lambda-\lambda_{\text{sol}}$ denotes the deviation of the squared length parameter from the classical solution.  We note that this quadratic action is monotonically increasing with growing $\lambda$, and that this behaviour is consistent with the asymptotic expression (\ref{time32}). 

\begin{figure}[t]
	\centering
	\includegraphics[width=.49\textwidth]{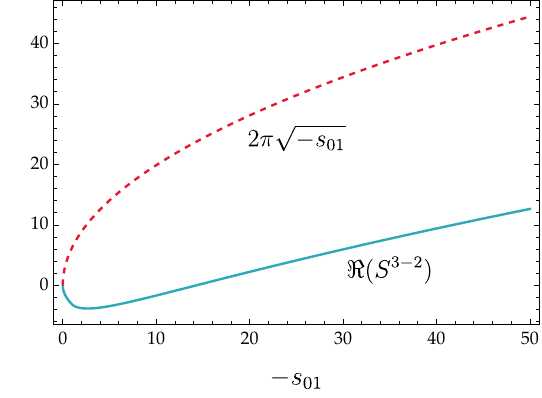}
	\caption{\label{Fig:32Example1-Comparison-Linearized}Exact Regge action $S^{3-2}$ (teal curve) in the neighborhood of the classical solution for the example configuration of boundary edges as in Fig.~\ref{Fig:32Example1} and asymptotic approximation to the Regge action (red dashed curve). Both curves grow monotonically for large bulk edge variable $s_{01}$, sufficiently far away from the flat solution. There is only one classical solution corresponding to the minimum of (the real part of) $S^{3-2}$.} 
\end{figure}

The linearized action describes the action around a solution, that is, for rather small bulk squared edge length. The asymptotic expression (\ref{time32}) describes the behaviour of the action for large edge length. Finding the same monotonic growth behaviour shows that we either have no further extrema of the action between the flat solution and the asymptotic regime, or there is an even number of such extrema.  For the specific case described in Fig.~\ref{Fig:32Example1}, one finds that there are indeed no further extrema between these two regimes, as can be seen from Fig.~\ref{Fig:32Example1-Comparison-Linearized}.

For the $4-1$ Pachner move we have four bulk variables. We assume that all of these are timelike and we gauge fix them all to be equal $s_{0i}=-\lambda$, with $\tilde \lambda =\lambda-\lambda_{\text{sol}}$ denoting the deviation from the flat solution. The quadratic action is then given by \cite{Borissova:2023izx}
\ba
S^{4-1}_{\text{qu}}  &=& - \frac{\tilde\lambda^2}{24} 
\frac{\sqrt{\abs{\mathbb{V}_{(1234)}}}
\sum_{1\leq i,j\leq 4} \sqrt{\abs{\mathbb{V}_{(\bar{i})}}} \sqrt{\abs{\mathbb{V}_{(\bar{j})}}} 
}
{
\sqrt{\abs{\mathbb{V}_{(0123)}}} \sqrt{\abs{\mathbb{V}_{(0124)}}} \sqrt{\abs{\mathbb{V}_{(0134)}}} \sqrt{\abs{\mathbb{V}_{(0234)}}}
}\, ,
\ea
where $(\bar{i})$ denotes the tetrahedron defined by the vertex set which is obtained by removing $(i)$ from $\{(0),(1),(2),(3),(4)\}$.

We note that this quadratic action is decreasing (if $\tilde \lambda$ is increasing), whereas the asymptotic expression (\ref{time41}) is increasing. One could therefore think that this is a case where there are  further extrema between the classical solution and the asymptotic case.

However, there is a different mechanism at work: We discussed below equation~(\ref{Vol3dA}) that a Lorentzian signature tetrahedron $(0123)$ with large timelike edges $(0i)$, $i=1,2,3$, needs to have a spacelike base triangle $(123)$. That is, for a Pachner move configuration $(01234)$ which admits large timelike edges $(0i)$, $i=1,2,3,4$, all the boundary triangles must be spacelike.

There are thus three possibilities for the signature of the tetrahedron $(1234)$: it is either Euclidean, Lorentzian or null. These cases are determined by the sign of the volume squared defined in (\ref{eq::VolumeCaleyMenger}).  We exclude  the non-generic null case and consider the Euclidean and the Lorentzian case separately. 
 
If the final tetrahedron is Euclidean, we do not have a classical solution. Instead, one has a complex saddle point, see~\cite{Asante:2021phx} for such an example.

If the final tetrahedron is Lorentzian, we can construct a family of classical solutions. But these classical solutions always include at least three spacelike edges. 

To see this, consider a Lorentzian tetrahedron whose triangles are all spacelike. Embed this tetrahedron into Minkowski space. Assume that this tetrahedron has at least one vertex, for example, vertex $(1)$, whose vertex angle contains a full light cone. (Since all triangles are spacelike, the vertex angle contains either no light cone or one full light cone.)  We furthermore choose the vertex $(4)$ and a point $i_{23}$ on  the edge $(23)$, and consider the triangle with vertices $(1),(4),i_{23}$. This triangle is either a Lorentzian triangle with only spacelike edges or a Euclidean triangle. If it is a Lorentzian triangle, then this triangle contains only one light cone at the vertex $(1)$. Thus, in both cases, the triangles do not contain light cones at the vertex $(4)$. This shows that there are no timelike directions emanating from $(4)$, but inside the tetrahedron $(1234)$. If we introduce a vertex $(0)$ inside this tetrahedron, the edge $(04)$ is therefore spacelike. This holds actually for all three edges $(0i)$ where $i=2,3,4$.

We thus see that for cases in which the boundary data in the $4-1$ configuration admits large timelike bulk edges, there does not exist a solution with only timelike bulk edges.

\section{Finite expectation values for spike and spine configurations}
\label{Sec:PathIntegralAsymptotics}

Spike and spine configurations which allow for infinitely large bulk edges, might lead to divergences for the quantum gravitational partition function. 

In Euclidean quantum gravity, spikes, such as those appearing in the $4-1$ configuration, are of particular concern. In two-dimensional Regge calculus, the work~\cite{Ambjorn:1997ub} showed that expectation values of sufficiently large powers of the lengths variables diverge due to spike configurations~\footnote{Here one assumes that the measure does not exponentially suppress large edge lengths.}.  In three- and four-dimensional Regge calculus, the $4-1$ and $5-1$ Pachner moves configurations, respectively, isolate the conformal factor degree of freedom of the spacetime metric~\cite{Dittrich:2011vz,Dittrich:2014rha}. The weights $\exp(-S_E)$ of Euclidean quantum gravity lead to an exponential enhancement of such configurations. In addition, there is an unbounded integration range, making Euclidean quantum Regge calculus highly problematic~\cite{Ambjorn:1997ub,Loll:1998aj}.

Here, we instead consider Lorentzian Regge calculus and investigate spine and spike configurations in the $3-2$ and $4-1$ Pachner move, respectively. We will consider the convergence properties of the path integral computing expectation values for powers of the  squared edge length.~\footnote{We note that the $4-1$ Pachner move, as discussed before, features a three-dimensional gauge symmetry~\cite{Dittrich:2008pw,Dittrich:2009fb}. The squared edge length is not invariant under this gauge symmetry. We nevertheless consider expectation values of the squared edge length, in order to compare with statements in previous literature~\cite{Ambjorn:1997ub}. To this end, we will apply a gauge fixing which can be also considered as a form of symmetry reduction. Namely, we will set all bulk edge lengths to be equal. We then compute the expectation value in this symmetry-reduced model.}
We will find that, although the integral is in general not absolutely convergent, one can extract finite expectation values. 

To define the Regge path integral, we have to specify the measure. Here, we allow for local measures which for large bulk variables scale with  a positive or negative (possible fractional) power of the absolute value of the bulk edge length squared. Let us note that, for three-dimensional Regge calculus, there is a preferred measure which to one-loop order guarantees triangulation invariance of the path integral~\cite{Dittrich:2011vz,Dittrich:2014rha, Borissova:2023izx}. For Lorentzian Regge calculus, this measure is given by 
\be\label{eq:Measure3d}
{\cal D}s_e\,=\,\mu(s_e)\prod_{e\subset \text{bulk}}\dd s_e\, =\, \frac{1}{\prod_{e\subset \text{bdry}}\sqrt{\sqrt{48}}}\frac{1}{\prod_{e\subset \text{bulk}}\sqrt{48}}\frac{\prod_\tau e^{-\imath \frac{\pi}{4}}}{\prod_{\tau} \abs{\mathbb{V}_\tau}^{\frac{1}{4}}}  \prod_{e\subset \text{bulk}}\dd s_e \,.
\ee

Here we aim to argue for the finiteness of expectation values. Therefore we only need to consider the asymptotic regime of large edge lengths. If the bulk edges are timelike, this regime is light-cone regular and the Regge action (ignoring boundary terms, which do not depend on the bulk lengths) is real. If, however, at least one of the bulk edges is spacelike, we have seen that the asymptotic regime is light-cone irregular, and the Regge action features an imaginary term. As discussed in Section~\ref{Sec:ReggeAction}, the sign for this imaginary term is ambiguous: the light-cone irregularities lead to a branch cut along the Lorentzian configurations, and the imaginary term changes sign across this branch cut~\cite{Asante:2021phx}. Clearly, the path integral will not converge if we decide to integrate along the side of the  branch cut where $\text{Im}(S)<0$. On the other hand, if we choose the opposite side of the branch cut, the imaginary part of the Regge action will lead to an exponential suppression of the amplitudes, and the path integral (as well as the integral for expectation values of powers of the edge length) converges. 

We therefore need to discuss only the light-cone regular asymptotic regimes, i.e., the cases where all bulk edges are timelike.

\subsection{$3-2$ Pachner move}

For the $3-2$ move, we have to integrate over one bulk edge. The asymptotic expression for the Regge action for a large timelike bulk edge with edge length squared $s_{01}=-\lambda$ is 
\ba
 S^{3-2} &=& 2\pi \sqrt{\lambda} + \,\, \mathcal{O}(\log \lambda)\,.
\ea
(As shown in equation~(\ref{eqGen32}), the same expression applies for the triangulation given by $N\geq 3$ tetrahedra sharing a timelike edge.)

\begin{figure}[t]
	\centering
	\includegraphics[width=.48\textwidth]{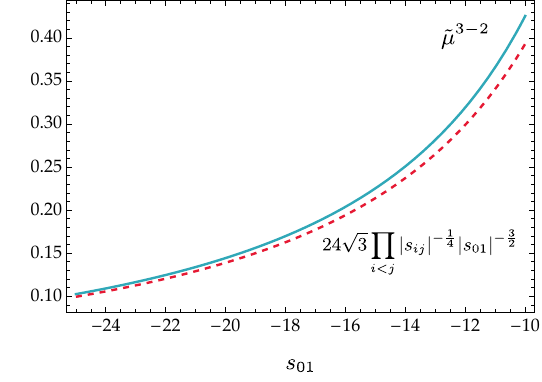}
	\caption{\label{Fig:32Measure} We show the exact expression for the measure for the $3-2$ Pachner, excluding the phase factors in equation~\eqref{eq:Measure3d} (teal curve), and its asymptotic behavior obtained by expanding the tetrahedral volumes according to equation~\eqref{eq:VolumeScaling3-2Move} in the limit of large timelike bulk edge $s_{01}$ (red dashed curve). The boundary data, compatible with the generalized inequalities, are chosen as ($s_{0i}=s_{1i}=1$, $s_{ij}=5$) with $i,j = 2,3,4$, $i<j$.}
\end{figure}

The measure factor in (\ref{eq:Measure3d}) in the asymptotic regime is proportional to
\ba
\mu^{3-2} \propto \lambda^{-3/2} \, .
\ea
Fig.~\ref{Fig:32Measure} compares the asymptotic behavior of the measure to the exact expression according to equation~\eqref{eq:Measure3d}. 
%For the curve representing the asymptotics, we have kept the boundary variables $s_{ij}$ which appear in the the expansions of the volumes for large timelike bulk edge, cf.~equation~\eqref{eq:VolumeScaling3-2Move}.

In order to consider the convergence behaviour for the expectation values of $s_{01}^n$, we use these asymptotic expressions and integrate $\lambda$ from a sufficiently large positive constant $c$ to $\infty$, 
\ba
\mathcal{I}_{3-2}(n,c) \,= \,\int_c^\infty \dd\lambda \, \, \lambda^{-3/2} \lambda^n \, e^{   2\pi \imath \sqrt{\lambda}  }\, .
\ea

Allowing for a more general measure given by some fractional positive or negative power $M$ of $\lambda$, we have to consider integrals of the type
\ba
\tilde{\mathcal{I}}_{3-2}(m,c)\,= \,
\int_c^\infty \dd\lambda \, \,  \lambda^m \, e^{   2\pi \imath \sqrt{\lambda}  } \,=\, 2\int_{\sqrt{c}}^\infty \dd\tilde{\lambda} \, \,  \tilde{\lambda}^{2m+1} \, e^{   2\pi \imath \tilde{\lambda} } \, . 
\ea
To evaluate these integrals, we introduce a regulator $\varepsilon>0$ and remove this regulator after performing the integral,
\ba\label{eq70}
\tilde{\mathcal{I}}_{3-2}(m,c) &=& 2\,\lim_{\epsilon\rightarrow 0}  \int_{\sqrt{c}}^\infty \dd\tilde{\lambda} \, \,  \tilde{\lambda}^{2m+1} \, e^{   (2\pi \imath -\varepsilon)\tilde{\lambda} } \nn\\
&=& 2 c^{m+1} E_{-2m-1}(-2 \pi \imath \sqrt{c}) \,,
\ea
where $E_n(z)=\int_1^\infty t^{-n} \exp(-zt)$ is the exponential integral function (analytically continued from the values of $n$ where it converges). $E_{-2m-1}(-2 \pi \imath \sqrt{c})$ is finite for $m \in \mathbb{R}$.

\begin{figure}[t]
	\centering
	\includegraphics[width=.48\textwidth]{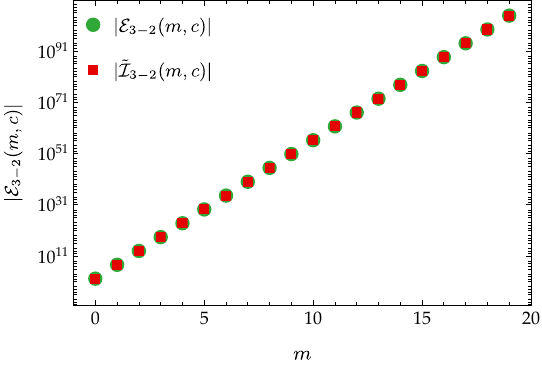}
	\caption{\label{Fig:32WynnAnalyt} Comparison of the integrals (\ref{eq70}) and (\ref{eq:32movefullPI}). The latter integral is  numerically evaluated using the Wynn algorithm, and we use for both integrals $c=250000$. We see that the integration over the full action is finite, and of the same order of magnitude as the approximated expectation value. Due to the choice of $c$, both results agree on a sub-percent level for any displayed value of $m$. %\BD{adjust notation in figure}
 }
\end{figure}

As a next step, we wish to evaluate the full expectation value for the light-cone regular cases. To that end, we neglect the measure term as it will effectively result in a shift of which expectation value is computed. We define 
\begin{equation}
\label{eq:32movefullPI}
{\mathcal{E}}_{3-2}(m,c) \,=\, \int_{\sqrt{c}}^{\infty} \dd \lambda  \lambda^{m} \, e^{\imath S^{3-2}(-\lambda)}\,,
\end{equation}
where $c$ is determined by the boundary data and the generalized triangle inequalities. 

While we cannot perform this integral analytically, we can employ series-acceleration methods such as Wynn's epsilon algorithm~\cite{Schmidt, Shanks, Wynn}, see also~\cite{WenigerReview} for a review.  The work~\cite{Dittrich:2023rcr} showed that Wynn's epsilon algorithm works well for path integrals (as well as for sums which appear in effective spin foam models~\cite{Asante:2020qpa, Asante:2021zzh}) and can also be employed to evaluate expectation values, even for cases where the underlying integral is not absolutely convergent.

In Fig.~\ref{Fig:32WynnAnalyt} we compare  the integral over the analytical approximation (\ref{eq70})  with the exact integral (\ref{eq:32movefullPI}). We used the same boundary configuration as for Fig.~\ref{Fig:32Example1}. To capture the asymptotic regime, we choose $c=250000$. Above this value the full action is approximated well by the asymptotic expression. Altogether, we see that the full result agrees with the asymptotic approximation on a sub-percent level.

We can thus conclude that the Regge expectation values of arbitrary powers of the squared bulk edge (and with a measure that asymptotically behaves as a fractional positive or negative power of the edge square) are finite. An essential mechanism to guarantee finiteness is the oscillatory nature of the Lorentzian path integral.

\subsection{$4-1$ Pachner move}

The  $4-1$ move configuration can be obtained through a subdivision of the final tetrahedron $(1234)$ by placing a vertex $(0)$ inside this tetrahedron, and by connecting all boundary vertices with the new inner vertex. Thus, if we consider the path integral for this configuration, we have to integrate over four bulk edges. As discussed previously, we consider only the case where all bulk edges are timelike. 

The $4-1$ move configuration comes with a three-parameter gauge symmetry \cite{Dittrich:2008pw,Dittrich:2009fb}: the solutions are flat, and can hence be constructed by embedding the final tetrahedron into flat space~\footnote{If this tetrahedron is demanded to satisfy the  triangle inequalities in some non-Lorentzian signature, one can construct a complex family of solutions in the same way.}, with the vertex $(0)$ placed at any point inside this tetrahedron~\footnote{If the path integral includes a sum over orientations of the tetrahedra, the vertex can be also placed outside the tetrahedron~\cite{Christodoulou:2012af, Dittrich:2014rha}}. This produces a three-parameter family of flat solutions.  Away from the flat solution, the gauge orbit is defined by demanding a constant action along the gauge orbits. As discussed previously,  we will consider the gauge fixing  that all the bulk edges have equal lengths.

The asymptotic expression for the Regge action for a $4-1$ configuration with four large timelike bulk edges of equal length square $s_{0i}=-\lambda$, is 
\ba
 S^{4-1} &=& 4\pi \sqrt{\lambda} + \,\, \mathcal{O}( \lambda^0)  \, .
\ea

The measure factor in (\ref{eq:Measure3d}) in the asymptotic regime is proportional to
\ba
\mu \propto \lambda^{-1} \,.
\ea
Fig.~\ref{Fig:32Measure} compares the asymptotic behavior of the measure to the exact expression according to equation~\eqref{eq:Measure3d}.  Using a gauge fixing $s_{01}=s_{02}=s_{03}=s_{04}$, we have to also insert a Faddeev-Popov determinant. Along the flat solution, where there is an explicit parametrization of the gauge orbits, this determinant can be straightforwardly computed to be (see~\cite{Baratin:2006yu,Dittrich:2011vz,Borissova:2023izx} for a similar computation)
\ba
{\cal F}=2^3\cdot 3!\,\, V_{(1234)}   \, ,
\ea
where $V_{(1234)}$ is the absolute volume of the final tetrahedron $(1234)$and is thus independent of the bulk edge lengths.

We are again interested in the convergence behaviour of the expectation values~\footnote{The inner edge lengths are not invariant under gauge transformations. But we use these expectation values to contrast with the results of~\cite{Ambjorn:1997ub} for Euclidean Quantum Regge calculus. The deficit angles are invariant under gauge transformations, at least in the linearized theory~\cite{Dittrich:2009fb,Bonzom:2013tna}. As we have shown in (\ref{eq:deficit41}), these behave asymptotically as $\mathcal{O}( \lambda^0)$. Assuming an asymptotic expansion in (negative) powers of $\lambda$ is possible, the convergence of expectation values of arbitrary powers of lengths implies the convergence of the expectation values of deficit angles.} of $\lambda^n$.  We also allow for a more general measure but demand that the measure, together with the Faddeev-Popov determinant, is (at least asymptotically) given by some fractional positive or negative power of $\lambda$. 

\begin{figure}[t]
	\centering
	\includegraphics[width=.48\textwidth]{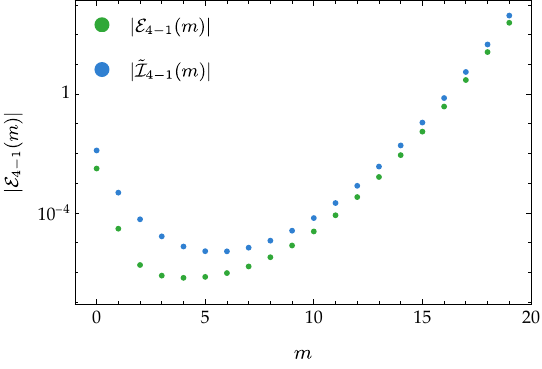}
	\caption{\label{Fig:41WynnAnalyt} Comparison of the approximated analytical expectation value of the squared edge length $\braket{\lambda^m}$ under the $4-1$ move with the numerical value extracted using the Wynn algorithm. We see that the integration over the full action is finite, and of the same order of magnitude as the approximated expectation value. This agreement becomes better for larger powers $m$.}
\end{figure}

Using the asymptotic approximations to investigate the convergence behaviour of the expectation values of $s_{0i}^n$, we have (absorbing measure factors into a shift of $n$ to $m$)
\ba
\tilde{\mathcal{I}}_{4-1}(m,c) \,= \int_c^\infty \dd\lambda \, \,  \lambda^m \, e^{   4\pi \imath \sqrt{\lambda}  }
\,=\, 2\int_{\sqrt{c}}^\infty \dd\tilde{\lambda} \, \,  \tilde{\lambda}^{2m+1} \, e^{   4\pi \imath \tilde{\lambda} } \, . 
\ea
In complete analogy to the $3-2$ move computation, these integrals evaluate to
\ba
\tilde{\mathcal{I}}_{4-1}(m,c) &=& 2 c^{m+1} E_{-2m-1}(-4 \pi \imath \sqrt{c}) \, .
\ea
We thus obtain a finite result for the expectation values of arbitrary powers of the bulk edge squares. 

Using the Wynn algorithm, we can also evaluate the expectation values  $\mathcal{E}_{4-1}$ using the full  Regge action.
In Fig.~\ref{Fig:41WynnAnalyt} we show the comparison of the results using the asymptotic approximation and the full Regge action for various values of $m$, and for the choice of $s_{0i}=-\lambda$, and $s_{ij}=5$, as well as $c=0$. The approximated result becomes more accurate for larger $m$, since with larger $m$, larger values of $\lambda$ dominate, where the asymptotic expression for the action provides a more accurate approximation.

\section{Discussion}\label{discussion}

In this paper, we investigate spike and spine configurations, which appear in the three-dimensional Lorentzian simplicical path integral for quantum gravity. They represent one of the difficulties to define a notion of scale in quantum gravity~\cite{Dittrich:2014ala,Asante:2022dnj,Carrozza:2016vsq}: although the bulk edges can be arbitrarily large, the volume of the boundary containing these arbitrarily large bulk edges can be very small, e.g., of Planck size.

Spikes and spines are therefore deeply quantum configurations and need to be studied to gain a better understanding of the quantum configuration space in Lorentzian quantum gravity.  The results of this paper include:
\begin{itemize}
\item The asymptotic behaviour of the action for the considered configurations is surprisingly simple: it is always linear in the length of the large bulk edge(s). The absolute value of the numerical coefficient is equal to $2\pi$ in the configuration where $N\geq 3$ tetrahedra share a bulk edge, and $4\pi$ in the $4-1$ Pachner move configuration.  
\item For the asymptotic behaviour of the action for the $4-1$ Pachner move configurations in the case where the signature of all bulk edges is the same, we found  dimensional reduction at work.  The Regge action does reduce to the Regge action of the two-dimensional boundary, which is a topological invariant. A similar dimensional reduction has been
observed for the three-dimensional Ponzano-Regge model and two-dimensional so-called intertwiner
models \cite{Dittrich:2013aia}, both of which are topologically invariant. Exploring such relationships might lead to new types of holographic behaviour in quantum geometric models \cite{Bonzom:2015ans,Dittrich:2017hnl,Dittrich:2017rvb,Dittrich:2018xuk}.
\item We have identified a simple method to decide which type of asymptotic regimes are allowed by the triangle inequalities. This allowed us to identify spine configurations, which appear in Lorentzian signature triangulations but are not allowed in Euclidean signature triangulations. Additionally, we have found that spike configurations, where the bulk edges adjacent to a vertex have different signatures, are allowed. 
\item On the other hand, we find that certain configurations with large bulk edges are forbidden by the Lorentzian triangle inequalities. E.g.~we cannot have a $4-1$ move configurations with spacelike boundary triangles and large spacelike bulk edges. Likewise, in four dimensions, a $5-1$ move configuration with spacelike boundary tetrahedra and large spacelike bulk edges is not allowed. But this is precisely the case, on which numerical efforts of the spin foam community have been concentrated: the latest result indicating that these configurations are finite (with a specific choice of measure)~\cite{Dona:2023myv}, is therefore not too surprising. There are many other types of $5-1$ Pachner move configurations, involving timelike tetrahedra, timelike triangles, or timelike edges where the triangle inequalities allow for large bulk edges. These configurations need to be investigated separately. 
\item We find many cases, in which the asymptotic regime leads to light-cone irregular configurations. To the best of our knowledge, this is the first time that such configurations have been discussed in the literature. In fact, all the configurations we considered here that involve at least one spacelike large bulk edge include light-cone irregularities.  This includes $4-1$ move configurations with only spacelike bulk edges,  as well as cases where the bulk edges can be timelike but include at least one large spacelike edge.
\item For all the light-cone irregular asymptotic regimes encountered in this paper, we find that they are predominantly of yarmulke type. In other words, there are bulk deficit angles which include less than two light cones. While for the $4-1$ configurations, we cannot rule out the presence of deficit angles of trouser type (that is, angles including more than two light cones), they are over-compensated by bulk deficit angles of yarmulke type.
\item  We find that the partition function and the expectation values for arbitrary powers of the bulk lengths are finite in the spine and spike configurations. This is in contrast to Euclidean Regge calculus, where the $4-1$ spike configurations isolate the conformal factor in the gravitational action~\cite{Dittrich:2011ke}, and (without rotating the sign of this factor) lead to divergences. We established this finiteness by using both the asymptotic expression for the Regge action with an analytical integration method and the exact expression for the Regge action with the Wynn algorithm~\cite{WenigerReview} to evaluate the integral.
\end{itemize}

The result on the finiteness of the partition function and expectation values shows how Lorentzian Quantum Regge calculus can circumvent one of the key problems of Euclidean Quantum Regge calculus. We note that we here do {\it not} need to employ a restriction on the signature of e.g.~the edges, as proposed (for two-dimensional triangulations) in~\cite{Tate:2011ct}. 

On the other hand, we have discovered many cases in which the asymptotic regime leads to light-cone irregular configurations. A crucial question for Lorentzian path integrals is how to deal with such light-cone irregular configurations. As far as we know, this is the first time light-cone irregular configurations have been identified, which extend over an infinitely large range of edge lengths. We discussed previously that along these light-cone irregular configurations, the Regge action acquires an imaginary term, the sign of which remains ambiguous due to branch cuts in the action. Due to the infinite range, there are only two choices: to either exclude such configurations or to choose the suppressing side of the branch cut for the integration contour. Indeed, a defining property of Causal Dynamical Triangulations~\cite{Loll:2019rdj} is that these configurations are forbidden to appear~\cite{Jordan:2013awa}. On the other hand, such light-cone irregular configurations, particularly those of yarmulke type, play a critical role for the derivation of entropy from the Lorentzian path integral~\cite{Marolf:2022ybi,Dittrich:2024awu}. In this case, one needs to choose the enhancing side of the branch cut for the integration contour~\cite{Dittrich:2024awu}.  This is allowed, as the branch cuts have only finite support. We believe that light-cone irregularities need to be better understood and might also need a finer classification to decide how to treat such configurations. One way to gain a better understanding of the geometry of the light-cone irregular configurations encountered here is to study the behaviour of null and timelike geodesics in such triangulations~\cite{Dittrich:2005sy}. 

The Lorentzian Ponzano-Regge model~\cite{Davids:2000kz,Freidel:2000uq} realizes a Lorentzian path integral for three-dimensional quantum gravity and is closely related to three-dimensional Lorentzian Quantum Regge calculus:  the amplitudes of this model asymptote to the cosine of the Regge action~\cite{Davids:2000kz}. A key difference is that the edge length squares in the Ponzano-Regge model are quantized, and the spectrum coincides with that of the $\text{SL}(2,\mathbb{R})$ Casimir operator. It would be very interesting to understand how light-cone irregular configurations appear in the Lorentzian Ponzano-Regge model. 

Another aspect which should be studied is the interplay of the asymptotics of the action we found here and the discrete spectrum for timelike edges in the Ponzano-Regge model, see the remarks in Section~\ref{Gen32}. These insights will be helpful to better understand four-dimensional spin foam models~\cite{Perez:2012wv,Engle:2007wy,Freidel:2007py,Baratin:2011hp,Asante:2020qpa,Asante:2021zzh}, as well as the appearance of potential divergences based on spike and spine configurations in this context.

%%%%%%%%%%%%%%%%%%%%%%%%%%%%%%%%%%%%%%%%%%%%%%

\appendix

\begin{acknowledgments}

BD thanks Jos\'e Padua-Arg\"uelles for discussions on related subjects. JB is supported by an NSERC grant awarded to BD and a doctoral scholarship by the German Academic Scholarship Foundation. DQ is partially supported by  the Blaumann Foundation.
Part of this research was conducted while BD was visiting the Okinawa Institute for Science and Technology  through the Theoretical Sciences Visiting Program. Research at Perimeter Institute is supported in part by the Government of Canada through the Department of Innovation, Science and Economic Development Canada and by the Province of Ontario through the Ministry of Colleges and Universities.

\end{acknowledgments}

%\bibliographystyle{plain}
%\bibliography{references}

\bibliographystyle{jhep}
\bibliography{references}

\end{document}